\pdfoutput=1
\documentclass{article}
\usepackage{caption} 
\usepackage{graphicx}
\usepackage{subfigure}
\usepackage{amsmath}
\usepackage{geometry}
\usepackage{authblk}
\usepackage{hyperref}
\usepackage[T1]{fontenc}
\usepackage[utf8]{inputenc}
\usepackage{float}
\usepackage{csquotes}

\usepackage{tikz}
	
\usetikzlibrary{backgrounds,fit,decorations.pathreplacing}

\title{A practical quantum encryption protocol with varying encryption configurations}
\author{Junxu Li, Zixuan Hu and Sabre Kais\thanks{Email: kais@purdue.edu}}
\affil{Department of Chemistry, Department of Physics and Astronomy, and

Purdue Quantum Science and Engineering Institute

Purdue University, West Lafayette, IN 47907, United States}

\begin{document}
\tikzset{meter/.append style={draw, inner sep=10, rectangle, font=\vphantom{A}, minimum width=30, line width=.8,
 path picture={\draw[black] ([shift={(.1,.3)}]path picture bounding box.south west) to[bend left=50] ([shift={(-.1,.3)}]path picture bounding box.south east);\draw[black,-latex] ([shift={(0,.1)}]path picture bounding box.south) -- ([shift={(.3,-.1)}]path picture bounding box.north);}}}
\tikzset{
 cross/.style={path picture={ 
\draw[thick,black](path picture bounding box.north) -- (path picture bounding box.south) (path picture bounding box.west) -- (path picture bounding box.east);
}},
crossx/.style={path picture={ 
\draw[thick,black,inner sep=0pt]
(path picture bounding box.south east) -- (path picture bounding box.north west) (path picture bounding box.south west) -- (path picture bounding box.north east);
}},
circlewc/.style={draw,circle,cross,minimum width=0.3 cm},
}
	\maketitle
	
	\begin{abstract}

	Quantum communication is an important application that derives from the burgeoning field of quantum information and quantum computation. Focusing on secure communication, quantum cryptography has two major directions of development, namely quantum key distribution and quantum encryption. In this work we propose a quantum encryption protocol that utilizes a quantum algorithm to create blocks of ciphertexts based on quantum states. The main feature of our quantum encryption protocol is that the encryption configuration of each block is determined by the previous blocks, such that additional security is provided. We then demonstrate our method by an example model encrypting the English alphabet, with numerical simulation results showing the large error rate of a mock attack by a potential adversary. With possible future improvements in mind, our quantum encryption protocol is a capable addition to the toolbox of quantum cryptography.	
	\end{abstract}
	
	\section*{Introduction}
	\label{introduction}

Utilizing quantum technologies for communication has been a major focus of the field of quantum computation and quantum information
\cite{divincenzo1995quantum, knill1998power,bennett2000quantum, barenco1995elementary}.
In particular, with emphasis on secure communication, quantum cryptography has seen enormous progress with both theoretical and experimental advances\cite{gisin2002quantum,ekert1991quantum,bennett1992quantum}.
One major direction of quantum cryptography, quantum key distribution (QKD)\cite{bennett1992quantum, jennewein2000quantum,xu2020secure, yin2020entanglement}, enables secure key generation and distribution by exploiting the non-locality of quantum entanglement. 
The other major direction of quantum cryptography, quantum encryption\cite{boykin2003optimal,hayden2004randomizing, hu2020quantum}, uses quantum computing techniques to create quantum states that carry the ciphertext.

The development of physical realizations of qubit systems and quantum circuits has led to a variety of breakthroughs including
the success of ground-to-satellite communication\cite{ren2017ground,liao2017satellite}, which enables reliable ultra-long-distance quantum key distribution;
and electron spin state teleportation with high fidelity, which proclaims the feasibility to achieve quantum teleportation in molecular systems\cite{rugg2019photodriven}.
There also arises pioneers of quantum teleportation in various systems such as atomic ensembles\cite{sherson2006quantum}, electron spins in quantum dots\cite{gao2013quantum}, trapped ions\cite{barrett2004deterministic} and superconducting circuits\cite{steffen2013deterministic}.
With these state-of-the-art advances in experimental techniques, one can envision the near-future realization of highly complex and sophisticated quantum communication protocols protected by quantum encryption methods.

In this work, we propose a quantum communication protocol with quantum encryption. As a block cipher, the plaintexts and the corresponding ciphertexts are sent in sequential blocks encrypted by a fixed number of qubits. The main feature that makes our quantum encryption method different from others is that the encryption configuration of each block is determined by the previous blocks, such that adjacent blocks are more likely to use different encryption configurations. This makes the encryption more difficult to break for a potential adversary. In the first section, we present the basic communication process in detail, where an example setup for the encryption and decryption processes is discussed. In the second section we discuss the security of the quantum encryption protocol against a potential adversary. In particular, we demonstrate the encryption of the English alphabet with an example model, showing the large error rate of a mock attack in a numerical simulation. In the last section we discuss potential improvements of the quantum encryption that aim to reduce noises and further increase security.

	\section{Theoretical framework}
	\label{Theoretical framework}

\begin{figure}[H]
	    \centering
        \begin{center}

        \includegraphics[width=0.75\linewidth]{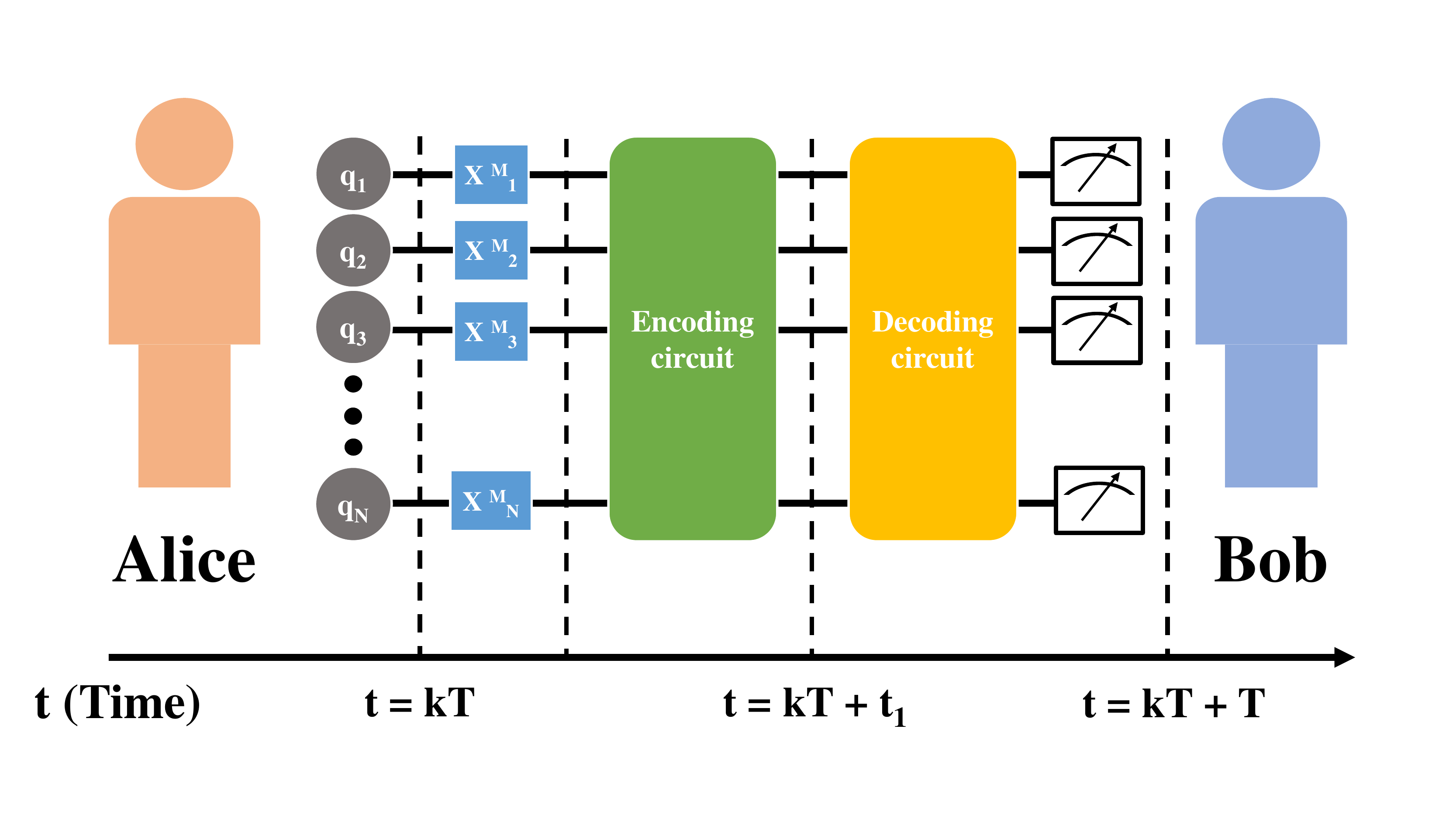}
        
	    \caption{{\bf Communication strategy between Alice and Bob.}
	    \\
	    Alice is isolated from the outside environment. 
	    Alice's friend, Bob attempts to communicate with Alice. The only connection between Alice and Bob is some shared qubits. Both Alice and Bob can apply arbitrary operations on the qubits, and Bob can apply $Z$ measurement on every single qubit.
	    When $t=kT$, where $k=0, 1, 2, 3\dots$, All qubits $q_1,q_2,\dots, q_N$ will be set at state $|0\rangle$. Alice will start preparing the qubits at state $|\Phi_k\rangle=|M_1M_2\dots M_N\rangle$, and $X$ represents Pauli-x operation. Later, these qubits will be encoded into state $|\Psi_k\rangle$ via the encoding circuit (the green box). Alice needs to finish these processes before $t=kT+t_1$, and $t_1<T$. Then Bob will start decode the qubits, and then apply Z measurements on each qubit. Measurements are required to be done before $t=kT+T$, after which all qubits will be reset at state $|0\rangle$.}
	    \label{Atime}
	    \end{center}
	\end{figure}

	Consider a scenario shown in fig.(\ref{Atime}), Alice is nearly isolated from the outside environment, and the only connection with outside environment is N qubits $\{q_1,q_2,\cdots,q_N\}$. 
	Alice attempts to communicate with her friend, Bob via these qubits. Alice works periodically: 
	{\bf 1.} At $t=kT, k=0,1,2,\cdots$, she will prepare $\{q_1,q_2,\cdots,q_N\}$ to be the state 
	$|\Psi_k\rangle$,
	where
	$k$ means the state is prepared at time $kT$. {\bf 2.} During $t\in(kT,kT+t_1]$, Bob can perform arbitrary operations and measurements on the qubits, while during $t\in(kT+t_1,kT+T]$, only Alice herself can operate on the qubits, where $0<t_1<T$, and both Alice and Bob have enough time to finish their operations. Noise is negligible. {\bf 3.} 
	They are still able to communicate even when Alice can not measure these qubits.
	When the noises and gate errors are ignored, we assume that Alice can not receive any information from outside.
	
	Alice and Bob apply a special quantum encoding circuit for encryption in their communication. 
	The plaintext is divided into blocks with the same length,
	The structure of the encoding circuit depends on the previous block plaintext.
	The $k-th$ block of the ciphertext $|\Psi_k\rangle$ can be described as a function of the initial plaintext $|\Phi_k\rangle$ and the encoding circuit $U_k$:
	\begin{equation}
	    |\Psi_k\rangle=U_k|\Phi_k\rangle
	    \label{eq1}
	\end{equation}
	where $U_k=U_k(|\Phi_{k-1}\rangle)$ is determined by the previous plaintext $|\Phi_{k-1}\rangle$.
	
\begin{figure}[H]
		\begin{center}
	\centerline{
    \includegraphics[width=0.75\linewidth]{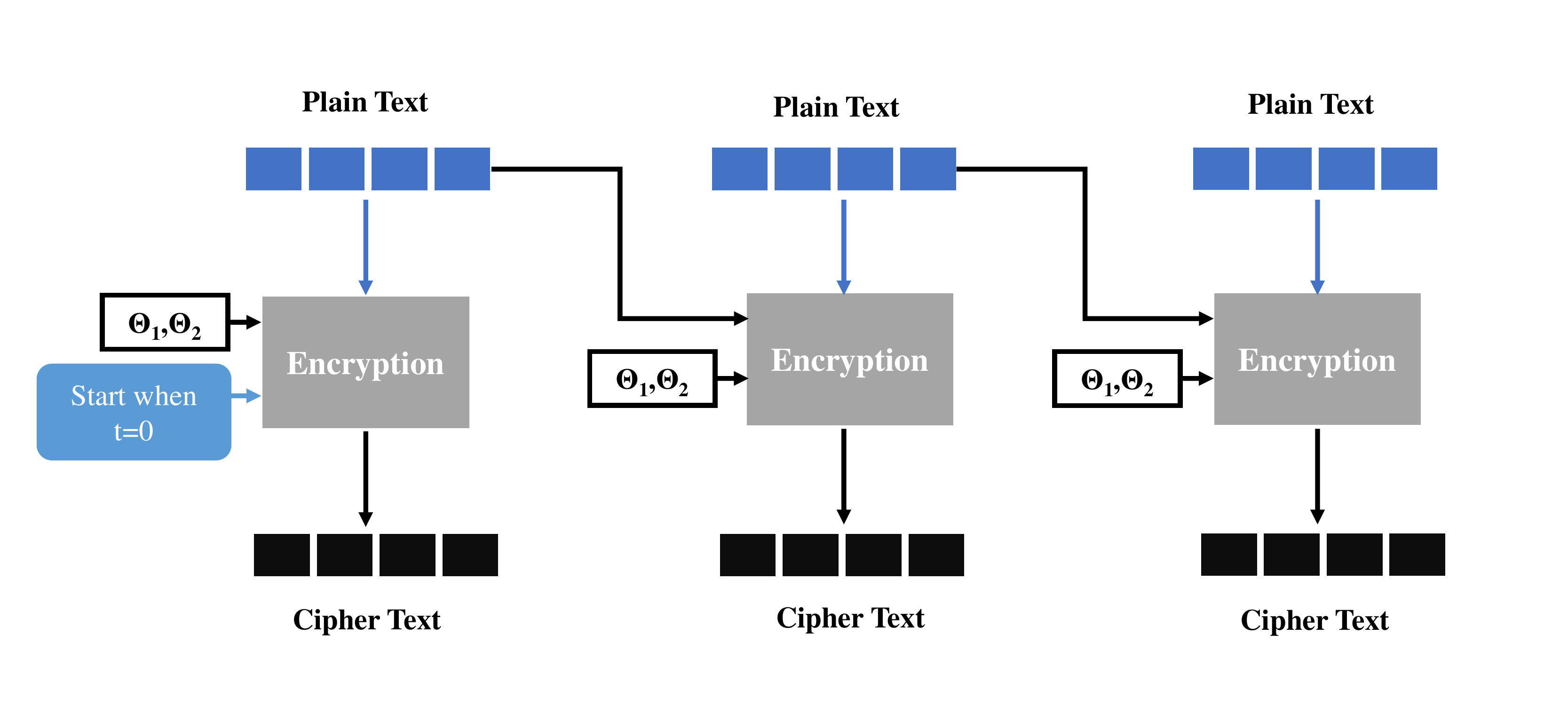}
    }
	\captionsetup{}
	\caption{
			{\bf Sketch of the encryption process.}\\	
				Binary message is divided into several blocks, each of them has the same length.
				Alice and Bob have set their clocks at the same time before being separated.
				They share the encoding operation encrypting the first block.
				When $k>1$, the $k-th$ ciphertext $|\Psi_k\rangle$ is generated by encrypting the plaintext $|\Phi_k\rangle$ and the encoding circuit $U_k$,
	            $|\Psi_k\rangle=U_k|\Phi_k\rangle$,
	            where the encoding operation $U_k=U_k(|\Phi_{k-1}\rangle)$ is determined by the former plaintext $|\Phi_{k-1}\rangle$ and $\Theta_{1,2}$.
	            $\Theta_{1,2}$ are some parameters shared by all encoding operations, will be discussed when we demonstrate the encoding operations.}
	\label{block_cipher}
	\end{center} 
    \end{figure}

	During each time interval T, Alice will transport N-bit information via the $N$ qubits. Note that one complete set of orthogonal eigenstates of these N qubits can be described as $\{|0\rangle,|1\rangle,\cdots,|2^N-1\rangle\}$, and an integer $n$ from 0 to $2^N-1$ can also represent the N-bit information just by rewriting $n$ into the binary digit form.
	At time $t=kT$, Alice attempts to transport message $n$ to Bob with the plaintext of the $k-th$ block $n(t=kT)$, and we denote $n(t=kT)$ as $n(k)$.
	For simplicity, in the following discussion the plaintext $|\Phi_k\rangle$ in eq.(\ref{eq1}) will be written as $|n(k)\rangle$. 
	Alice encodes information into the quantum state 
	$|\Psi_k\rangle=U_k|n(k)\rangle$,
	where $U_k=U_k(n(k-1))$ is determined by the former $n$ instead of a constant operation.
	They make an agreement that 
	the first block is encrypted by the encoding circuit $U_0=U_0(|0\rangle)$.
	Fig.(\ref{block_cipher}) is a sketch of the encryption process.
	When $k>1$, the $k-th$ ciphertext $|\Psi_k\rangle$ is generated by encrypting the plaintext $|n(k)\rangle$ and the encoding circuit $U_k$,
	$|\Psi_k\rangle=U_k|n(k)\rangle$,
	where the encoding operation $U_k=U_k(|n({k-1})\rangle)$ is determined by the former plaintext $|n({k-1})\rangle$ and $\Theta_{1,2}$.
	$\Theta_{1,2}$ are some parameters shared by all encoding operations, and will be discussed when we demonstrate the encoding operations.
	
	Generally, it requires a number of repeating measurements to get a quite accurate estimation of a certain quantum state, as an example, the widely used quantum tomography\cite{d2001quantum} requires exponential measurements to pin a quantum state. In addition, Alice's special communication strategy makes it extremely difficult to extract the information from the N qubits.
	As no copy is offered, Bob needs to ensure that $n(k)$ can be derived after only one single measurement (Here we ignored the noises and errors, strategies against noises in the circuit are discussed later). 
	Besides, the encoding operator $U_k$ is determined by the former plaintext $n$, so that all following results can no longer be convincing if we make a wrong estimation for even one single $n$.
	However, next we show that knowing the first encoding operator and the way how the following encoding operators depend on the previous plaintexts, concequently Bob can indeed obtain n(k) with ease. The difficulty created by the communication strategy therefore falls to Eve, the potential adversary that does not know the encoding operators.
	
\begin{figure}[H]
	    \centering
	    \subfigure[]{
        \centering
        \includegraphics[width=0.56\linewidth]{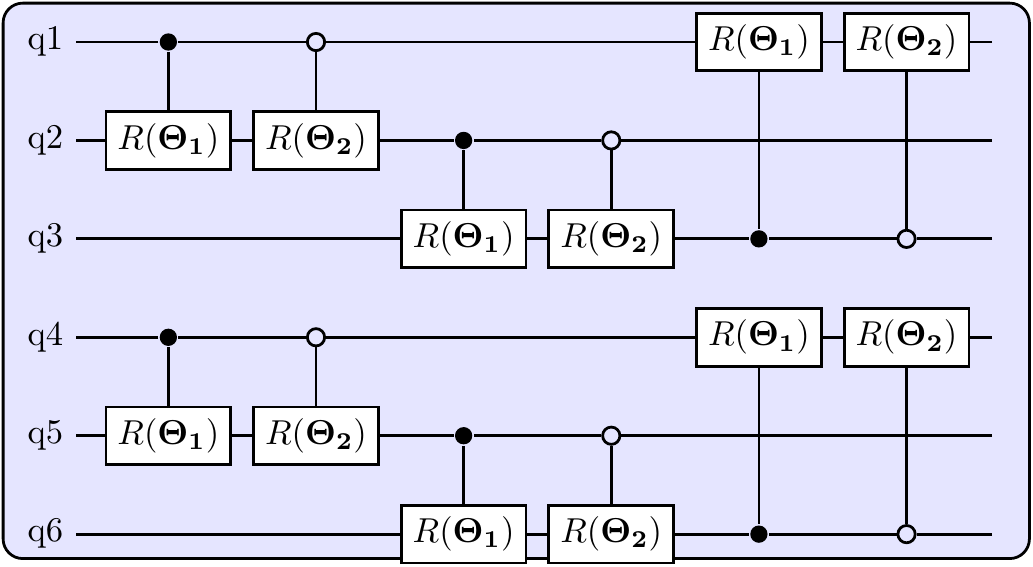}
        \label{triangle}
        }
        \subfigure[]{
        \centering
        \includegraphics[width=0.4\linewidth]{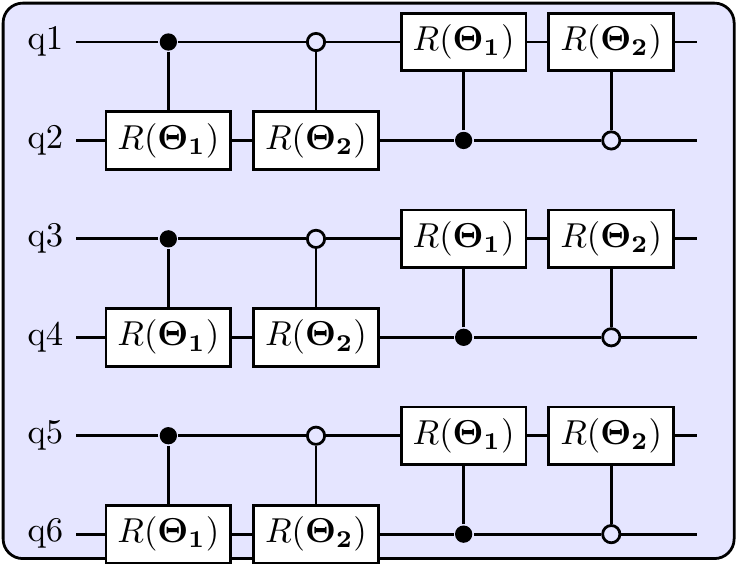}
        \label{2qubit}
        }
	    \caption{{\bf Sketch of encoding circuit.}\\
	            Initially, $q_1,q_2,\cdots,q_6$ are prepared at state $|n\rangle$, then Alice can use these circuit to encode information into $|\Psi_k\rangle$.
				(a.)The encoding circuit based on 3 qubits control gates loop. ${q_1,q_2,q_3}$ form one loop and ${q_4,q_5,q_6}$ form another.  We note this encoding circuit as $U_{tri}$, and one can use $U_{tri}^{-1}$ to extract information encoded by $U_{tri}$.
				(b.)The encoding circuit based on 2 qubits control gates loop. $q_1,q_2$ form one loop, and $q_3,q_4;q_5,q_6$ form two other loops respectively. We would like to note this circuit as $U_{bin}$, and one can apply $U_{bin}^{-1}$ as well to extract information encoded by $U_{bin}$.
				Here we only show two simple encryption operations as example.
				There are in fact more choices, which will be discussed later.
				}
	    \end{figure}

	Here, we will demonstrate the whole process of encryption and decryption. For simplicity, we assume that 6 qubits $\{q_1,q_2,\cdots,q_6\}$ are used by Alice to communicate with Bob.
	Generally, 10 numbers, 26 capital letters and 26 small ones, a mark to divide words (blank space) and a mark to divide sentence (like $,$ or $.$) are required in communication, so that in total 64 states  $|\Psi\rangle$ are needed. As $2^6=64$, 6 qubits are already enough to encode the English alphabet together with numbers and marks. Sometimes special characters might be required in the communication, and methods to design encoding circuits for more qubits are presented in the supplementary materials.
	In the following example model, all $U_k$ can be decomposed into 2-qubit control gates $u_{ij}$
	\begin{equation}
	    u_{ij}=|0\rangle\langle0|_i\otimes R_j({\bf \Theta_1})+ |1\rangle\langle 1|_i\otimes R_j({\bf \Theta_2})
	    \label{uij}
	\end{equation}
	where $i,j\in\{1,2,3,4,5,6\},i\neq j$. $q_j$ is controlled by $q_i$. $R({\bf \Theta})$ is single qubit rotation gate, and ${\bf \Theta}=(\theta_1,\theta_2,\theta_3,\theta_4)$ is a four dimensional vector, $R({\bf \Theta})=\exp{(-i\theta_1)R_z(\theta_2)R_y(\theta_3)R_z(\theta_4)}$. For example, in the simplest situation, $U_k$ has only two possible choices, and these two encoding circuit are as following: 
	
	{\bf Encoding circuit 1: 3-qubit-loop.} ${q_1,q_2,q_3}$ form one loop of control rotation gates and ${q_4,q_5,q_6}$ form another. Initially, $q_1,q_2,\cdots,q_6$ are prepared at state $|n\rangle$, and $n=0, 1, 2, \dots, 2^N-1$, where there are $N$ qubits used in their communication. Then Alice can use the circuit shown in fig.(\ref{triangle}) to encode information into $|\Psi_k^{in}\rangle$. We note this circuit as $U_{tri}$, and one can use $U_{tri}^{-1}$ to extract information encoded by $U_{tri}$.
	
    {\bf Encoding circuit 2: 2-qubit-loop.} As shown in fig.(\ref{2qubit}), $q_1,q_2$ form one loop, and $q_3,q_4;q_5,q_6$ form two other loops respectively. Initially, $q_1,q_2,\cdots,q_6$ are prepared at state $|n\rangle$, then Alice can use this circuit to encode information into $|\Psi_k^{in}\rangle$. We would like to note this circuit as $U_{bin}$, and one can apply $U_{bin}^{-1}$ as well to extract information encoded by $U_{bin}$.
	
\begin{figure}[H]
		\begin{center}
	\centerline{
    \includegraphics[width=0.96\linewidth]{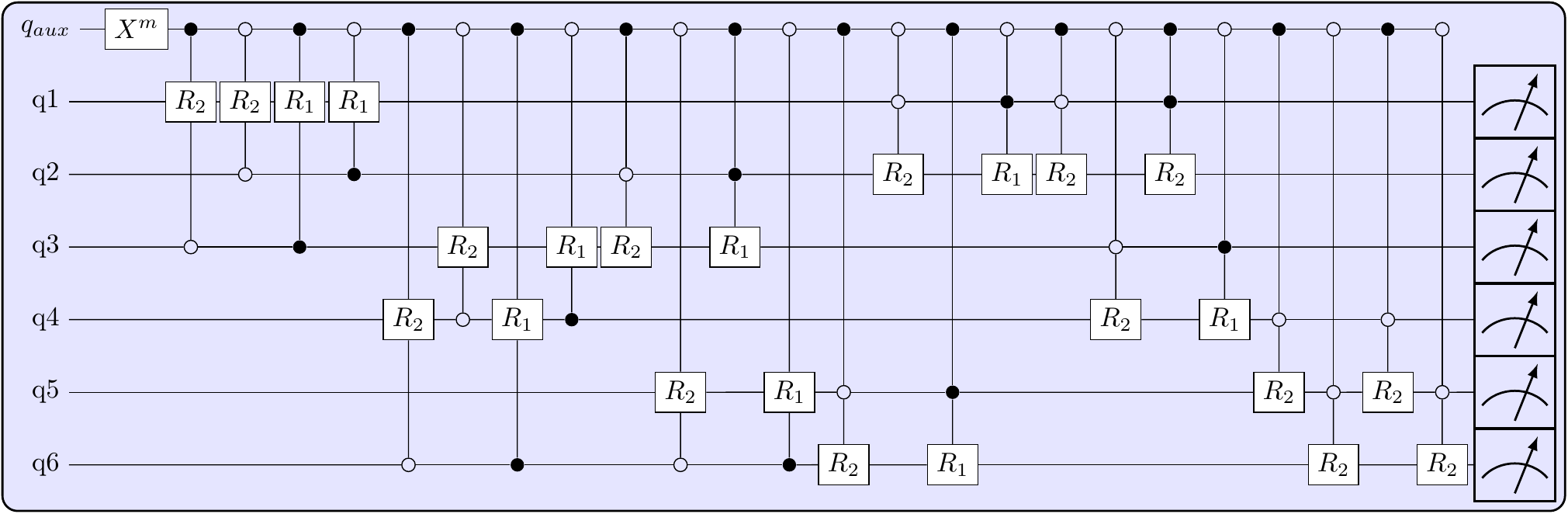}
    }
	\captionsetup{justification=centering}
	\caption{
			{\bf A possible circuit to decode Alice's information}\\	
				$q_{aux}$ is an auxiliary qubit, and in each period it is preset at state $|0\rangle$.
				$m$ is the previous measurement results of $q_6$.
				As Alice and Bob make an agreement that the first block is encrypted by $U_0=U_0(|0\rangle)$, Bob will set $m=0$ initially.
				Alice only performs operations on $q_1~q_6$.
				For simplicity, here we use $R_1$ and $R_2$ to represent $R^{-1}({\bf \Theta_1})$ and $R^{-1}({\bf \Theta_2})$ respectively.}
	\label{detect}
	\end{center} 
\end{figure}

	If $n(k-1)$ is odd, then Alice will use circuit 1 to encode $n(k)$; otherwise, she will choose circuit 2 to encode $n(k)$. Obviously, the encoding strategy is determined by the bit stored in $q_6$, as when $q_6$ represents 1 then Alice will use circuit 1 for encoding, otherwise circuit 2 will be chosen. 
    Bob knows the parameters ${\bf \Theta_1, \Theta_2}$, then as shown in fig.(\ref{detect}), he can combine the inverse of circuit 1 and 2 together as the decoding circuit, where $R_1$ and $R_2$ represent $R^{-1}({\bf \Theta_1})$ and $R^{-1}({\bf \Theta_2})$ respectively. One auxiliary qubit $q_{aux}$ is introduced to represent the previous measurement result to $q_6$. In fact, such a circuit is not the only possible solution to decode information from Alice's special communication strategy. If we prefer to make it more difficult to decode, it is also a choice to design more different encoding operations for every state $|n\rangle$, yet more auxiliary qubits will be required for such complicated strategies. 
	
    Note that even in the simplest examples displayed above, we still apply control rotation gates instead of standalone single qubit gates.
    The existence of multi qubit gates in encryption ensures that the change of a single qubit in the plaintext could have influence on more than one qubit in the cipher, which corresponds to the concept of diffusion in classical cryptography\cite{shannon2001mathematical}.

	\section{Application in quantum communication}
	\label{Application in quantum communication}
	
\begin{figure}[H]
	    \centering
        \includegraphics[width=0.75\linewidth]{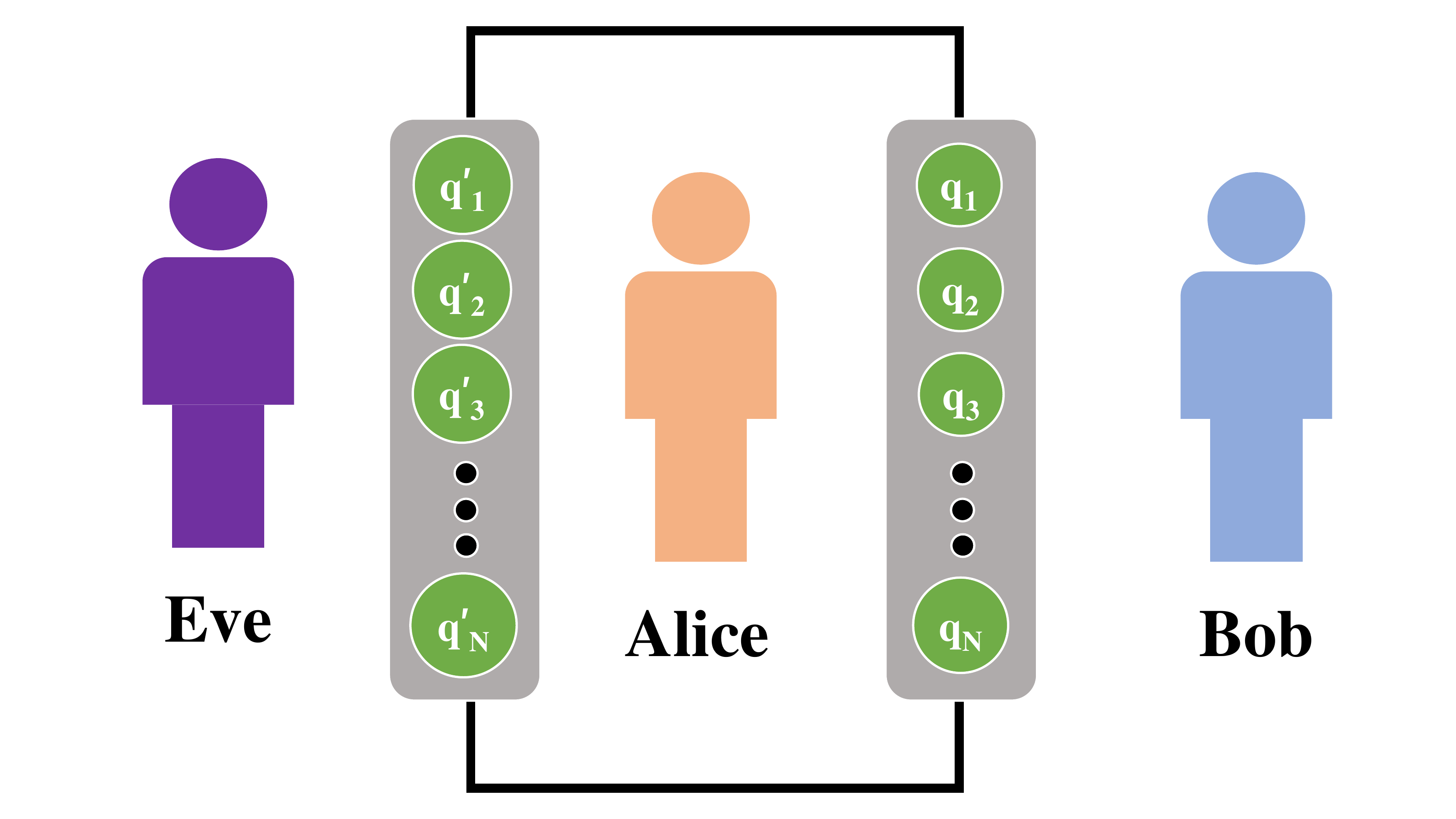}
	    \caption{{\bf Sketch of the relation among Alice, Bob and Eve}
	    Eve(left) attempts to wiretap the communication. Alice(mid) is locked in a 'black room', and the only connection between Alice and the outside environment is some qubits. Alice attempts to communicate with her friend Bob(right) via the qubits. However, there are two groups of qubits: $q_1, q_2, \dots, q_N$ and $q'_1, q'_2, \dots, q'_N$. Bob has access to operate and measure $q_1, q_2, \dots, q_N$, while Eve has the access to apply operation or measurement on $q'_1, q'_2, \dots, q'_N$. However, Alice does not know which group is under Bob's control. Consequently, Alice has to prepare two groups as same state.
	    }
	    \label{ABC}
	\end{figure}

	In this section, we will study a more complicated situation. 
	Assume that Eve attempts to eavesdrop the communication between Alice and Bob. As shown in fig.(\ref{ABC}), Still we design a scenario where Alice is nearly isolated from the outside environment, and the only connection between Alice and people outside is some qubits. Alice attempts to communicate with Bob, her friend via these qubits. However, to wiretap the communication, Eve Eve has prepared her own group of qubits to impersonate Bob. Now there are two groups of qubits: $q_1, q_2, \dots, q_N$ and $q'_1, q'_2, \dots, q'_N$. Bob has access to operate and measure $q_1, q_2, \dots, q_N$, while Eve has the access to apply operation or measurement on $q'_1, q'_2, \dots, q'_N$. However, Alice does not know which group is under Bob's control. Consequently, she has to prepare two identical groups of ciphertext states.
	
	In section I, we have used two encoding operations: $U_{tri}=(u_{31}\cdot u_{23} \cdot u_{12})\otimes (u_{64}\cdot u_{56} \cdot u_{45})$ (the 3-qubit encoding operations, shown in fig.(\ref{triangle})), and $U_{bi}=(u_{21}\cdot u_{12})\otimes(u_{43}\cdot u_{34})\otimes(u_{65}\cdot u_{56})$ (the 2-qubit encoding operations, shown in fig.(\ref{2qubit})), where $u_{ij}$ is described by eq.(\ref{uij}).
	And we use $U_{tri}$ to encode the new state $|n(k)\rangle$ if the previous $n(k-1)$ is odd, and use $U_{bi}$ otherwise.
	Generally speaking, if Eve does not know the encoding strategy, or she can not apply operations on the qubits, then it will be quite safe for Alice and Bob to communicate via this strategy (For more details see the first section in the supplementary materials).
	
	Review the information shared between Alice and Bob as following.
	
	{\bf S1.} Alice will start to prepare qubits at certain states for communication since $t=0$. Before $t=0$ she will produce some random state. Alice and Bob have set their clock at same time before they are separated.
	
	{\bf S2.} Six qubits are used in the communication. Note the eigenstates under $Z$ measurements can be written as $|n\rangle$, where $n$ is an integer from 0 to 63. State $|0\rangle$ represents the blank space, used as word divider. $|1\rangle$ to $|26\rangle$ represent capital letters 'A' to 'Z', $|27\rangle$ to $|52\rangle$ represent lower-case letters 'a' to 'z', and $|53\rangle$ to $|62\rangle$ represent numbers '0' to '9'. The last eigenstate $|63\rangle$ represents ',', '.', or other marks to divide sentences.
	
	{\bf S3.} Alice will prepare state $|\Psi(k)\rangle=U(n(k-1))|n\rangle$, where $U(n(k-1))=U_{tri}$ if $n(k-1)$ is odd and $U(n(k-1))=U_{bi}$ if $n(k-1)$ is even. Alice and Bob set that $U(t=0)=U_{bi}$.
	
	Even if we assume that Eve knows the general strategy that the encoding operation of each block is determined by the plaintext of the previous block, without knowing the particular selection of encoding operations from the total set, and without knowing the plaintext of the previous block, the chance of her guessing the correct decoding operation and obtaining the plaintext is very low.
	
	Then consider the worst situation where Eve also knows the selection of encryption operations and parameters $\Theta_{1,2}$. 
	Besides, we assume that Eve has the authority to apply arbitrary operations on the qubits $q'_1, q'_2, \dots, q'_N$. 
	In other words, Eve knows almost all information that Alice and Bob share, with only one exception, the clock, which obstructs Eve from applying correct decryption operation on the first block. 
	In order to miss the least information, by large chance Eve will start wiretapping as early as possible. 
	According to the first communication strategy, before $t=0$, Alice produces states randomly. On the other hand, Eve has no idea about the exact '$t=0$'.
	Consequently, at $t=0$, Eve might apply the decoding operation $U_{bi}$, or $U_{tri}$. Once Eve applies $U_{tri}$, she can hardly decode the first state $|\Psi(t=0)\rangle$ correctly, then she will get less information comparing to Bob.
	
	Though we have shown that there is possibility to avoid Alice from getting all information, we need to note that it is nearly impossible to prohibit Eve from getting 'much information' under such communication strategies. Even though Eve might decode state $|n\rangle$ as $|n'\rangle$ by mistake, once both $n, n'$ are odd or even, the following states will all be decoded correctly.
	
	Take the above consideration into account, Alice and Bob can change the third strategy as the following:
	
	{\bf S3.}Alice will prepare state $|\Psi(k)\rangle=U(n(k-1))|n\rangle$.
	Instead of having only two encoding options determined by the parity of the previous plaintext, Alice and Bob can pick up a total number of n different encoding options, determined by all n different possibilities of the previous block of plaintext.
	All $U(n)$ can be decomposed as combination of $u_{ij}=u_{ij}({\bf \Theta_1}, {\bf \Theta_2})$ described by eq.(\ref{uij}).
	
	In the supplementary materials, we will provide one encoding operations set as example for $S.3$, which is also used in the following numerical simulations. For instance, Alice attempts to transport the famous quotation of Alexandre Dumas,
	
	\begin{quote}
	    All human wisdom is contained in these words: Wait and hope.
	    \\
	    \rightline{The Count of Monte Cristo, Chap 117.\qquad\qquad}
	\end{quote}

	As in the encoding process, all marks such as $',.:'$ will be recognized as $'.'$, and since there is no character representing 'line break' or 'new line', the quotation will be converted into,
	
	\begin{quote}
	    All human wisdom is contained in these words. Wait and hope. The Count of Monte Cristo. Chap 117.
	\end{quote}
	
	Alice and Bob arrange that the $|\Psi(t=0)\rangle$ is encoded by $V_0$, and encoding operation $V_m$ will be applied for the $k-th$ block $t$ if $n(k-1)=m$. For simplicity, definition of the encoding operations are presented in the supplementary materials.
	At $t=0$(according to Alice and Bob's clock), Alice will start to prepare the 6 qubits at state
	\begin{equation}
	    |\Psi(t=0)\rangle
	    =
	    V_0|1\rangle
	\end{equation}
	where $|1\rangle$ is the plaintext, representing the first character $'A'$ in the quotation. 
	Before $t=t_1$, Alice should finish the preparation, and then Bob will start his decoding process. 
	Bob will apply decoding operation $V_0^{-1}$, as he already knows that $|\Psi(t=0)\rangle$ is encoded by $V_0$. For simplicity, here we ignore the noise and errors in the whole process (New rules shall be introduced into the communication strategy against noise, and we will give a brief discussion later). After $Z$ measurement on each qubit, and Bob will derive that $|\Phi(t=0)\rangle=|1\rangle$. Bob finishes his measurement before $t=T$, and at $t=T$ all qubits are reset at state $|0\rangle$. The next character is $'l'$, corresponding to state $|38\rangle$ (or $|100110\rangle$ as binary). Now Alice should prepare state
	\begin{equation}
	    |\Psi(t=T)\rangle=V_1|38\rangle=V_1|100110\rangle
	\end{equation}
	Alice and Bob can keep communicating in this way. Ignoring the errors and noise, Bob can always receive the correct information theoretically.
	
\begin{figure}[H]
	    \centering

	    \subfigure[]{
        \includegraphics[width=0.45\linewidth]{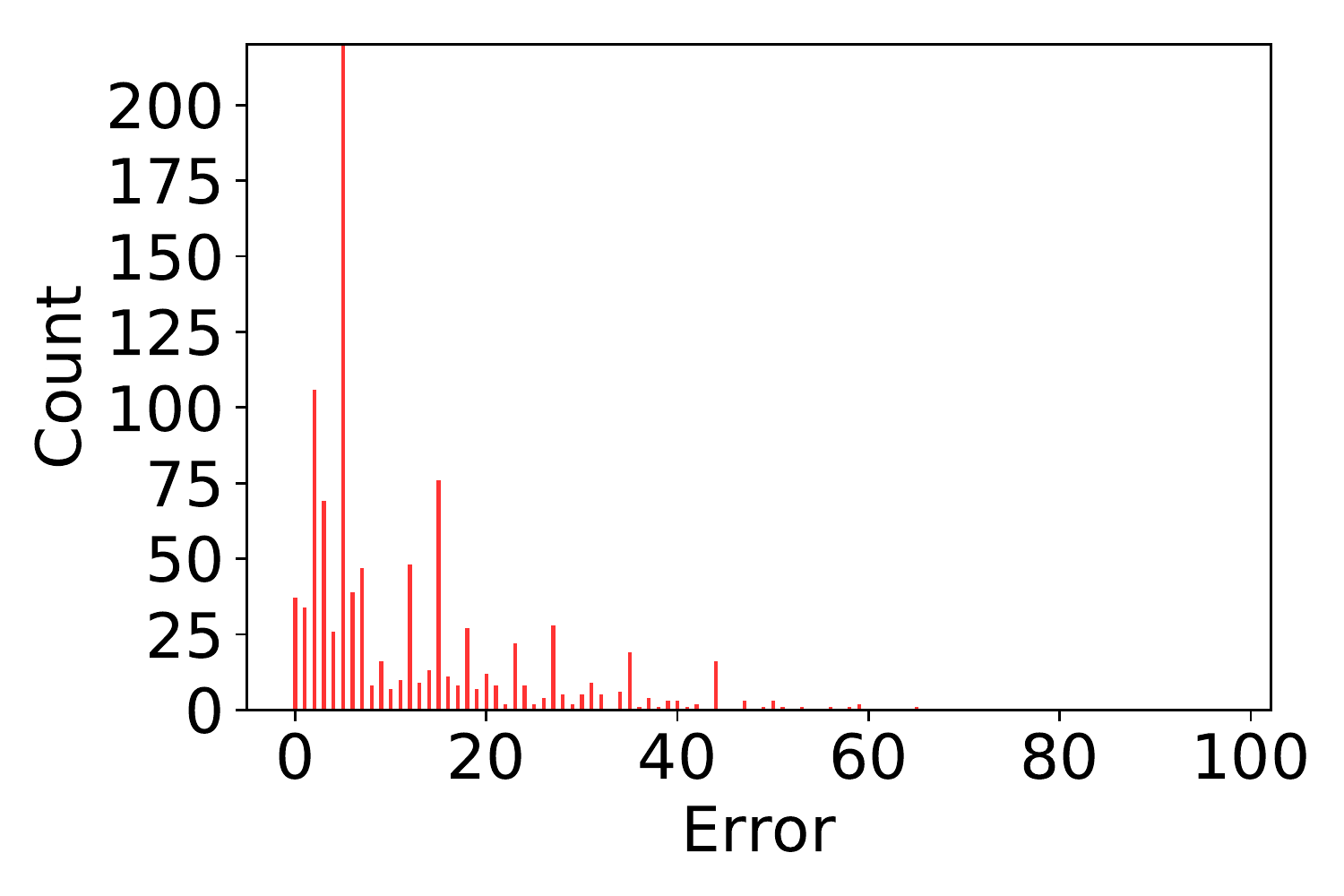}
        \label{64en1pre}
        }
        \subfigure[]{
        \includegraphics[width=0.45\linewidth]{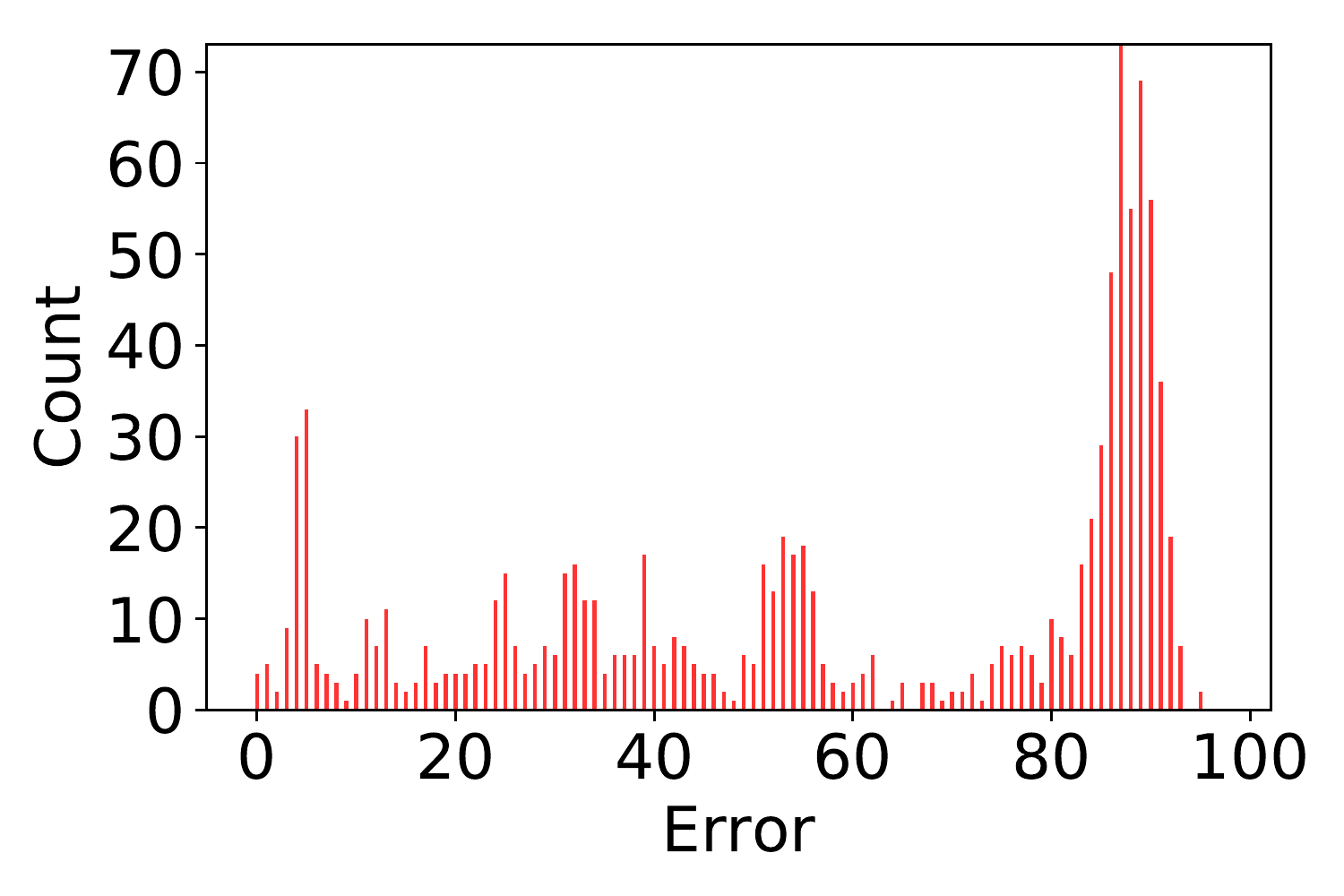}
        \label{64en2pre}
        }

	    \caption{{\bf Numerical simulation for Eve's decoding process.}
	    \\
	    Assume that Eve starts decoding before $t=0$, and at $t=0$ she has uniform probability to apply each decoding operation. 'Err' represents the total errs that Eve makes when decoding the sentence. 'Count' represents the frequency of certain total mistakes. A rectangle that locates at $Error=x$ with $Count=y$ infers that in the whole simulation, there are $y$ times that Eve makes $x$ times mistakes. 1000 times simulation are carried in total, $\Theta_1=(0, 0.15\pi, 0.72\pi, 0.32\pi)$, $\Theta_2=(0, 0.45\pi, 0.17\pi, 1.64\pi)$.
	    (a.) Only one previous state has contribution to the encoding operation. (b.)Two previous states are included to decide the next encoding operation.
	    }
	    
	\end{figure}

	On the other hand, Eve finds it impossible to find $t=0$ accurately for the absence of shared clock.
	Additionally, Alice have already noticed her existence and will produce random states before $t=0$. 
	Here we can safely assume that Eve applies random decoding operations at $t=0$. 
	Assume that Alice has no bias when generating the random states before $t<0$, Eve applies each decoding operation with the same possibility, $1/64$. Fig.(\ref{64en1pre}) shows the numerical simulation results of Eve's decoding process when the encoding operation of each block is determined by only the previous plaintext. 
	There are 97 characters in the sentence. From the histogram, one may notice that by a quite large chance, Eve will decode the whole sentence with fewer than 20 total mistakes. 
	Obviously now the communication strategy is not safe enough, when the encoding operation of each block is determined by only one previous plaintext.
	In the following section, we will provide improvement to increase the security in communication.
	
	To improve the security against wiretapping, consider the following encoding operation,
	\begin{equation}
	    U(k) = V[(n(k-1)+n(k-2))mod64]
	    \label{2pre}
	\end{equation}
	where operation $V$ are still the 64 encrytion operations presented in the supplementary materials.
	Now the encoding operation at time $t$ is decided by the two former blocks $n(t-T)$ and $n(t-2T)$. 
	In our example, at $t=0$, Alice attempts to send the letter $'A'$, $n(t=0)=1$, and at $t=T$ Alice attempts to send $'l'$, $n(t=T)=38$. $(1+38)mod64=39$, so Alice will encode the third character (still $'l'$) as
	\begin{equation}
	    |\Psi(t=2T)\rangle=V_{39}|38\rangle
	\end{equation}
	Still, Alice and Bob make an agreement that the first block is encrypted by $V_0$.
	Even though Alice will generate states randomly before $t=0$, Bob does not need to measure them, and these random states will not effect their communication. In fig.(\ref{64en2pre}), we show the numerical simulation results of Eve's decoding process under the improved communication strategy. Now we Eve can find it extremely difficult to decode the communication. 
	She might decode one character accidentally, yet which does little help for the following block, unless she can decode two characters simultaneously. 
	Comparing with fig.(\ref{64en1pre}), there is greater chances for Eve to make more mistakes, and the information transfer is safer than using only one block to determine the encoding operation. 

	Note that here we only introduced the former two states into encoding operation. Theoretically, one can expand eq.(\ref{2pre}) into
	\begin{equation}
	    U(t) = V\left[(\sum_{\tau=t-t'}^{t-1}n(\tau))mod64\right]
	    \label{npre}
	\end{equation}
	Hence, one can include arbitrary number of previous states into the determination of the encoding process. 
	
	Till now, the key in our design contains three parts.
	The first one is the selection of the encoding configurations from all possible configurations that can be applied on the plaintexts to create ciphertexts. These define the controls and targets of the control-unitary gates, and examples of the encoding configurations can be found in Figure 3 and the supplementary materials.
	Second part of the key contains the parameters $\Theta_{1,2}$.
	These define the actual actions of the control-unitary gates and small differences difference of them will lead to extremely different encrytion operations.
	In this paper only the simplest situation is discussed, where $\Theta_{1,2}$ are used multi times in the encryption circuit.
	To introduce more potential choices for the key, one can replace the repeated $\Theta_{1,2}$ as independent parameters.
	Generally the more parameters there are in the encryption circuits, there will be more potential choices for the key and the communication will be more safe.
	The last part is the ways that the encoding configurations are selected by the plaintext of the previous blocks.
	Numerical simulation shows that the communication is still reliable even when the second and third parts of the key is released.
	More theoretical discussions about quantum encryption can be found in our recent work\cite{hu2020quantum}.
	
	Noise and errors are all ignored in the above discussion, while the noise might lead to some mistakes in real experiments. 
	Here we provide a 4th strategy against noise and errors with error-detecting auxiliary characters:
	
	{\bf S4.} 
	At time $t=klT$, where $k=1, 2\dots$, and $l$ is a constant positive integer, Alice will prepare the qubits at some certain states, which depend on previous states.
	After decoding state $|\Psi(t=klT)\rangle$, Bob will make out if he made any mistakes during $[(k-1)lT, klT]$. 
	To make up the mistake, they need one more wire that Bob is able to use it to send messages back to Alice. 
	Once Bob find the state $|\Psi(t=klT)\rangle$ out of expectation, he can send Alice a message, and then Alice will restart communication from the state $|\Psi(t=(k-1)lT+T)\rangle$.
	
	Strategy {\bf 4} works on the plaintexts, instead of the encoding process. Here we would provide one example to demonstrate how it works. To make it possible for Bob to self check the encoding process, Alice rewrites the message by adding one auxiliary character every 9 characters (So that auxiliary characters will be the $10k-th$ character). She sets the auxiliary character at position $10k$ to be the same value as the $[10(k-1)-1]-th$ one. The message now will be
	
	\begin{quote}
	    All human{\bf' '} wisdom i{\bf'n'}s contain{\bf'i'}ed in the{\bf'n'}se words.{\bf'e'} Wait and{\bf'.'} hope. Th{\bf'd'}e Count o{\bf'h'}f Monte C{\bf'o'}risto. Ch{\bf'C'}ap 117.
	\end{quote}
	
	Characters embraced by apostrophes are auxiliary characters (Please note that the apostrophes are included to mark the auxiliary characters, but they themselves are not part of the message). Here Alice and Bob set the first auxiliary character as blank space. 
	The second auxiliary character is 'n', same as the $9^{th}$ character.
	The third auxiliary character is 'i', same as the $19^{th}$ character.
	Other auxiliary characters are generated similarly.

	\section{Conclusion}
	\label{conclusion}
	
	In this work we have proposed a protocol of quantum encryption  with varying encryption configurations. 
	The plaintext is divided into blocks with the same length, and represented by the eigenstates under Z-measurements. 
	Operation encrypting each single block is determined by one or more previous blocks of the plaintexts.
	Thus, successful decryption of the former block is required when attempting to extract information from a new one.
	Key in the protocol contains several parts, the selection of configurations of the encryption operations, the parameters that determine the encryption operations and the ways the encoding operations are determined by the previous plaintext blocks.
	All parts are essential for successful encryption and decryption.
	Further, we studied the protection against wiretapping.
	Simulation results show that it is still difficult  to decode the communication between Alice and Bob even when part of the key is released.
	Finally, we also discussed an error-correction method against noises in the encryption and decryption circuits.
	
	\section*{Acknowledgment}
	\label{acknowledgment}
	We acknowledge the financial support by Purdue Research Foundation and funding by the U.S. Department of Energy (Office of Basic Energy Sciences) under Award No.de-sc0019215.

	\bibliography{ref}

\begin{thebibliography}{10}

\bibitem{divincenzo1995quantum}
David~P DiVincenzo.
\newblock Quantum computation.
\newblock {\em Science}, 270(5234):255--261, 1995.

\bibitem{knill1998power}
Emanuel Knill and Raymond Laflamme.
\newblock Power of one bit of quantum information.
\newblock {\em Physical Review Letters}, 81(25):5672, 1998.

\bibitem{bennett2000quantum}
Charles~H Bennett and David~P DiVincenzo.
\newblock Quantum information and computation.
\newblock {\em nature}, 404(6775):247--255, 2000.

\bibitem{barenco1995elementary}
Adriano Barenco, Charles~H Bennett, Richard Cleve, David~P DiVincenzo, Norman
  Margolus, Peter Shor, Tycho Sleator, John~A Smolin, and Harald Weinfurter.
\newblock Elementary gates for quantum computation.
\newblock {\em Physical review A}, 52(5):3457, 1995.

\bibitem{gisin2002quantum}
Nicolas Gisin, Gr{\'e}goire Ribordy, Wolfgang Tittel, and Hugo Zbinden.
\newblock Quantum cryptography.
\newblock {\em Reviews of modern physics}, 74(1):145, 2002.

\bibitem{ekert1991quantum}
Artur~K Ekert.
\newblock Quantum cryptography based on bell’s theorem.
\newblock {\em Physical review letters}, 67(6):661, 1991.

\bibitem{bennett1992quantum}
Charles~H Bennett, Gilles Brassard, and N~David Mermin.
\newblock Quantum cryptography without bell’s theorem.
\newblock {\em Physical review letters}, 68(5):557, 1992.

\bibitem{jennewein2000quantum}
Thomas Jennewein, Christoph Simon, Gregor Weihs, Harald Weinfurter, and Anton
  Zeilinger.
\newblock Quantum cryptography with entangled photons.
\newblock {\em Physical Review Letters}, 84(20):4729, 2000.

\bibitem{xu2020secure}
Feihu Xu, Xiongfeng Ma, Qiang Zhang, Hoi-Kwong Lo, and Jian-Wei Pan.
\newblock Secure quantum key distribution with realistic devices.
\newblock {\em Reviews of Modern Physics}, 92(2):025002, 2020.

\bibitem{yin2020entanglement}
Juan Yin, Yu-Huai Li, Sheng-Kai Liao, Meng Yang, Yuan Cao, Liang Zhang, Ji-Gang
  Ren, Wen-Qi Cai, Wei-Yue Liu, Shuang-Lin Li, et~al.
\newblock Entanglement-based secure quantum cryptography over 1,120 kilometres.
\newblock {\em Nature}, 582(7813):501--505, 2020.

\bibitem{boykin2003optimal}
P~Oscar Boykin and Vwani Roychowdhury.
\newblock Optimal encryption of quantum bits.
\newblock {\em Physical review A}, 67(4):042317, 2003.

\bibitem{hayden2004randomizing}
Patrick Hayden, Debbie Leung, Peter~W Shor, and Andreas Winter.
\newblock Randomizing quantum states: Constructions and applications.
\newblock {\em Communications in Mathematical Physics}, 250(2):371--391, 2004.

\bibitem{hu2020quantum}
Zixuan Hu and Sabre Kais.
\newblock A quantum encryption scheme featuring confusion, diffusion, and mode
  of operation.
\newblock {\em arXiv preprint arXiv:2010.03062}, 2020.

\bibitem{ren2017ground}
Ji-Gang Ren, Ping Xu, Hai-Lin Yong, Liang Zhang, Sheng-Kai Liao, Juan Yin,
  Wei-Yue Liu, Wen-Qi Cai, Meng Yang, Li~Li, et~al.
\newblock Ground-to-satellite quantum teleportation.
\newblock {\em Nature}, 549(7670):70, 2017.

\bibitem{liao2017satellite}
Sheng-Kai Liao, Wen-Qi Cai, Wei-Yue Liu, Liang Zhang, Yang Li, Ji-Gang Ren,
  Juan Yin, Qi~Shen, Yuan Cao, Zheng-Ping Li, et~al.
\newblock Satellite-to-ground quantum key distribution.
\newblock {\em Nature}, 549(7670):43--47, 2017.

\bibitem{rugg2019photodriven}
Brandon~K Rugg, Matthew~D Krzyaniak, Brian~T Phelan, Mark~A Ratner, Ryan~M
  Young, and Michael~R Wasielewski.
\newblock Photodriven quantum teleportation of an electron spin state in a
  covalent donor--acceptor--radical system.
\newblock {\em Nature chemistry}, 11(11):981--986, 2019.

\bibitem{sherson2006quantum}
Jacob~F Sherson, Hanna Krauter, Rasmus~K Olsson, Brian Julsgaard, Klemens
  Hammerer, Ignacio Cirac, and Eugene~S Polzik.
\newblock Quantum teleportation between light and matter.
\newblock {\em Nature}, 443(7111):557--560, 2006.

\bibitem{gao2013quantum}
WB~Gao, Parisa Fallahi, Emre Togan, Aymeric Delteil, YS~Chin, Javier
  Miguel-Sanchez, and A~Imamo{\u{g}}lu.
\newblock Quantum teleportation from a propagating photon to a solid-state spin
  qubit.
\newblock {\em Nature communications}, 4(1):1--8, 2013.

\bibitem{barrett2004deterministic}
MD~Barrett, J~Chiaverini, T~Schaetz, J~Britton, WM~Itano, JD~Jost, E~Knill,
  C~Langer, D~Leibfried, R~Ozeri, et~al.
\newblock Deterministic quantum teleportation of atomic qubits.
\newblock {\em Nature}, 429(6993):737--739, 2004.

\bibitem{steffen2013deterministic}
Lars Steffen, Yves Salathe, Markus Oppliger, Philipp Kurpiers, Matthias Baur,
  Christian Lang, Christopher Eichler, Gabriel Puebla-Hellmann, Arkady Fedorov,
  and Andreas Wallraff.
\newblock Deterministic quantum teleportation with feed-forward in a solid
  state system.
\newblock {\em Nature}, 500(7462):319--322, 2013.

\bibitem{d2001quantum}
GM~D'Ariano and P~Lo Presti.
\newblock Quantum tomography for measuring experimentally the matrix elements
  of an arbitrary quantum operation.
\newblock {\em Physical review letters}, 86(19):4195, 2001.

\bibitem{shannon2001mathematical}
Claude~Elwood Shannon.
\newblock A mathematical theory of communication.
\newblock {\em ACM SIGMOBILE mobile computing and communications review},
  5(1):3--55, 2001.

\end{thebibliography}

    \newpage
    \section*{Supplementary Materials}
    \subsection*{Optimization of parameters-When Eve can only measure qubits}
    In this section we will offer one way to optimise the 6-qubit model for communication application by adjusting parameters $\Theta_1, \Theta_2$. 
    For simplicity, here Alice encodes the qubits only with operations shown in fig.(3,4).
    On one hand, our circuit should be robust, as we need to make sure that Bob, who is to receive information via the qubits, can still make correct decision when the quantum gates are imperfect or when there are environmental noises. In real experiments, the control gates can not always be perfect as we designed. When one sets a $R_z(\theta)$ rotation gate, often a $R_z(\theta+\Delta\theta)$ gate is set, and generally we have $|\Delta\theta|\ll\theta$. On the other hand, we need to make sure that Eve, who attempts to wiretap the communication, can not get too much information. Further, if we know that Alice has little chance to get any information after $t=N_AT$, then we can use the first $N_A$ quantum states to transport some trivial information, so that Eve can get nothing useful. Here, we note $P_B(k|{\bf \Theta_1,\Theta_2})$ as the probability of Bob to get all information correctly before $t=kT$, and $P_A(k|{\bf \Theta_1,\Theta_2})$ for Alice. Aim of the optimisation process is to find suitable parameters ${\bf \Theta_1,\Theta_2}$ so that for a given constant $c>1$, there exists an integer $N_T>0$, $P_B(cN_T|{\bf \Theta_1,\Theta_2})$ is significantly larger than $P_A(N_T|{\bf \Theta_1,\Theta_2})$, after which we can ensure that the 6-qubit model can be used to transport $6(c-1)N_T$ bits information during time period $N_TT$.
	
	At $t=0$, the initial quantum state is $|0\rangle$, so that $|\Psi^{in}_1\rangle$ is certainly encoded by the 2-qubit loop shown in fig.(\ref{2qubit}). Consider the minimum unit of 2-qubit loop, qubits $q_{1}$ and $q_{2}$. After the encoding process they will be prepared at state
	\begin{equation}
	    |\psi_{1,2}\rangle=U(0)|q_1\rangle\otimes\langle 0|U(q_1)|q_2\rangle|0\rangle + U(1)|q_1\rangle\otimes\langle 1|U(q_1)|q_2\rangle|1\rangle
	    \label{state}
	\end{equation}
	For simplicity, here we defined the unitary operator $U(q), q=0,1$ and $U(0)=R({\bf \Theta_2})$ and $U(1)=R({\bf \Theta_1})$. Assume Eve chooses measurement $M_1$ on $q_1$ and $M_2$ on $q_2$, and their corresponding eigenstates are $|\phi_1^+\rangle,|\phi_1^-\rangle$ and $|\phi_2^+\rangle,|\phi_2^-\rangle$, where $|\phi^+\rangle=M|0\rangle$ and $|\phi^-\rangle=M|1\rangle$. For chosen measurements, we can rewrite eq.(\ref{state}) as
	\begin{equation}
	    \begin{split}
	        |\psi_{1,2}(q_1,q_2)\rangle
	        &=\left[
	        \langle\phi_1^+|R(0)|q_1\rangle\langle0|R(q_1)|q_2\rangle\langle\phi_2^+|0\rangle+
	        \langle\phi_1^+|R(1)|q_1\rangle\langle1|R(q_1)|q_2\rangle\langle\phi_2^+|1\rangle
	        \right]|\phi_1^+\phi_2^+\rangle\\
	         &+\left[
	        \langle\phi_1^-|R(0)|q_1\rangle\langle0|R(q_1)|q_2\rangle\langle\phi_2^+|0\rangle+
	        \langle\phi_1^-|R(1)|q_1\rangle\langle1|R(q_1)|q_2\rangle\langle\phi_2^+|1\rangle
	        \right]|\phi_1^-\phi_2^+\rangle\\
	         &+\left[
	        \langle\phi_1^+|R(0)|q_1\rangle\langle0|R(q_1)|q_2\rangle\langle\phi_2^-|0\rangle+
	        \langle\phi_1^+|R(1)|q_1\rangle\langle1|R(q_1)|q_2\rangle\langle\phi_2^-|1\rangle
	        \right]|\phi_1^+\phi_2^-\rangle\\
	         &+\left[
	        \langle\phi_1^-|R(0)|q_1\rangle\langle0|R(q_1)|q_2\rangle\langle\phi_2^-|0\rangle+
	        \langle\phi_1^-|R(1)|q_1\rangle\langle1|R(q_1)|q_2\rangle\langle\phi_2^-|1\rangle
	        \right]|\phi_1^-\phi_2^-\rangle
	    \end{split}
	\end{equation}
	Then we can define a new function $f(M_1,M_2|{\bf \Theta_1,\Theta_2})$ to describe how good the measurements $M_1,M_2$ is for Eve, and
	\begin{equation}
	    f(M_1,M_2|{\bf \Theta_1,\Theta_2})
	    =\sum_{q_1',q_2'=0}^1
	    {\left[
	    {|\langle\phi_1(q_1')\phi_2(q_2')|\psi_{1,2}(q_1',q_2')\rangle|}^2
	    -\sum_{q_1'',q_2''\neq q_1',q_2'}^{1}{|\langle\phi_1(q_1')\phi_2(q_2')|\psi_{1,2}(q_1'',q_2'')\rangle|}^2
	    \right]}
	\end{equation}
	where we note $\phi_{1,2}(1)=\phi^+_{1,2}$ and $\phi_{1,2}(0)=\phi^-_{1,2}$. Theoretically, for given $\Theta_1,\Theta_2$, the maximum of $f(M_1,M_2|{\bf \Theta_1,\Theta_2})$ only depends on $M_1,M_2$, so that we can define function $g(\Theta_1,\Theta_2)$ as the maximum result. By adjusting $\Theta_1,\Theta_2$, Alice and Bob can decide the optimal parameters that ensure least possibility for Eve to derive correct information.
	
	In fact, if Eve can only measure each qubits instead of applying arbitrary operations on the qubits, then Alice and Bob do not need to spend time optimizing the parameters.
	
\begin{figure}[H]
	    \centering
        \includegraphics[width=0.75\linewidth]{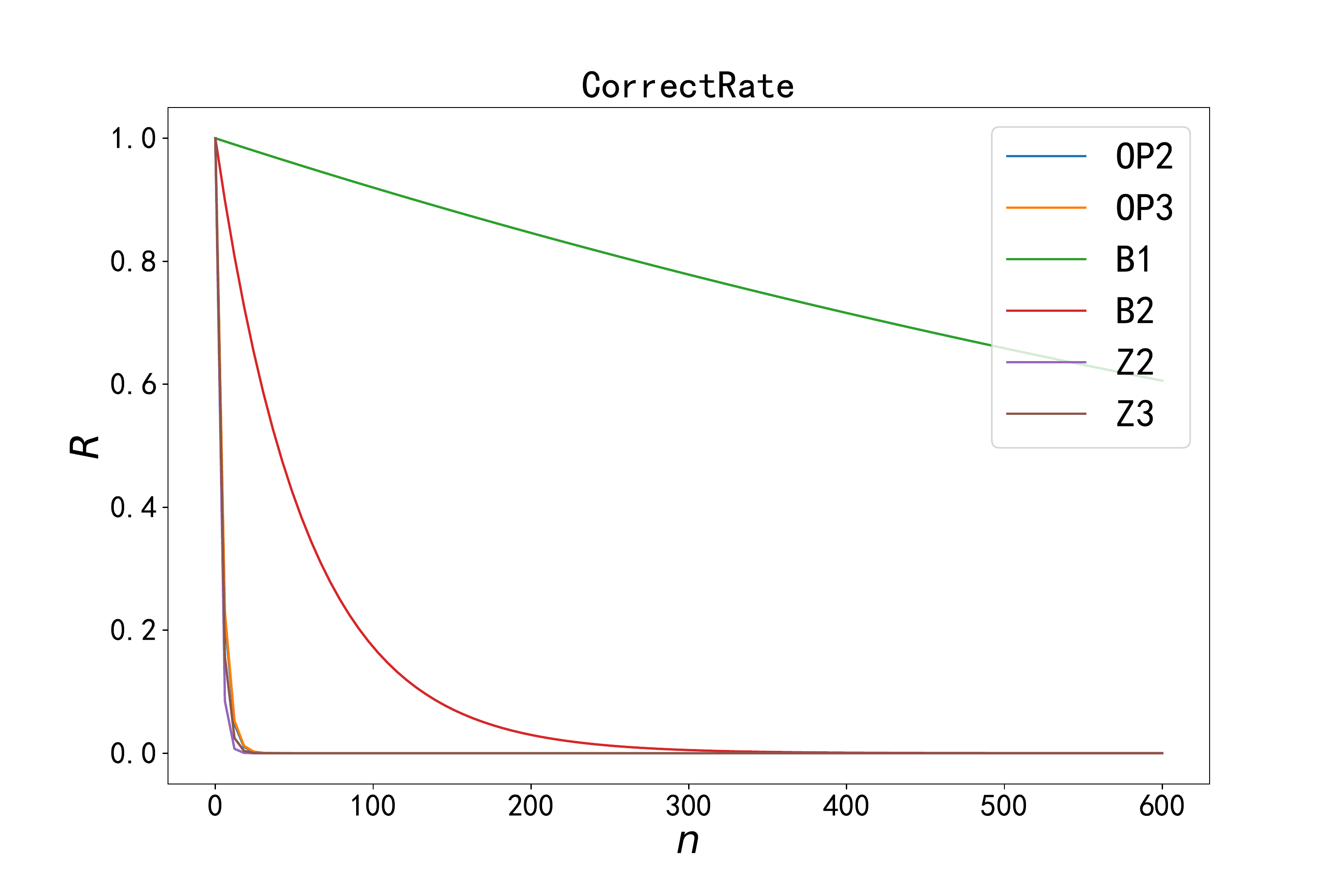}
	    \caption{{\bf Simulation result of correct rate for different decoding strategies.} Here, we simulate the correct rate $R$ for different decoding strategies when transporting $n$ bits information. {Z2:} Assume that all information are encoded by circuit fig.(\ref{2qubit}), and one applies only $Z$ measurement on every qubit. {Z3:} Assume that all information are encoded by circuit fig.(\ref{triangle}), and one applies only $Z$ measurement on every qubit. {OP2:} Assume that all information are encoded by circuit fig.(\ref{2qubit}), and one applies the optimal measurement on every single qubit. {OP3:} Assume that all information are encoded by circuit fig.(\ref{triangle}), and one applies the optimal measurement on every single qubit. {B1:} One can apply the decoding circuit fig.(\ref{detect}). Due to the environment noise, fidelity of all operations is 0.995. {B2:} One can apply the decoding circuit fig.(\ref{detect}). Due to the environment noise, fidelity of all operations is 0.9.}
	    \label{matchrate}
	\end{figure}

	Here, we simulate the correct rate for different decoding strategies, as shown in fig.(\ref{matchrate}). We assume that the document is transported via binary numbers, and every time Alice will transport 6 bit information via the 6 qubits. Qubits are encoded by circuit fig.(\ref{2qubit}) if the former number is odd, or by circuit fig.(\ref{triangle}) if the former number is even. $R$ represents the correct rate, or the possibility to derive all information correctly when Alice transports totally $n$ bits information. 
	$Z2$(and $Z3$) represents that Eve directly apply $Z$ measurements on every single qubit, and assume that the first state is encoded by the encoded by circuit fig.(\ref{2qubit}) (fig.(\ref{triangle})).
	$OP2$(and $OP3$) represents that Eve optimized her measurements on every single qubit, and assume that the first state is encoded by the encoded by circuit fig.(\ref{2qubit}) (fig.(\ref{triangle})).
	$B1$(and $B2$) shows the performance of Bob's decoding process with different noise, the self-check strategy is not introduced.
	One can find that if only measurements on single qubit are allowed, it is nearly impossible to wiretap the communication, even though there are only two encoding operations. In the simulation, we set ${\bf \Theta_1}=(0.45\pi, 4.04, 1.04, 0.92)$ and ${\bf \Theta_2} = [0, 0.35, 0.55\pi, 0.79)$(Just some random numbers, not the optimal ones).

    \subsection*{Find the encoding operation from a single state}
    Here we will provide another method to find the encoding operation from one single state, and only consider the encoding operations shown in fig.(\ref{triangle}) and fig.(\ref{2qubit}) .
    
    Assume that the first few states $|\Psi(t=0,1,2,\cdots, k)\rangle$ are encoded by one same encoding circuit, which is either circuit 1 or circuit 2. All details of these two circuits are already known, then it is also possible to find the encoding method from $|\Psi(t=0,1,2,\cdots, k)\rangle$. Here we will demonstrate the basic operations. For simplicity, note that the 2 possible encoding methods as $U_1$, $U_2$, and one complete set of this system as $\{|0\rangle,|1\rangle,|2\rangle,\cdots ,|2^N-1\rangle\}$. Then we have
    \begin{equation}
    |\psi_{1,n}\rangle=U_1|n\rangle,\qquad
    |\psi_{2,n}\rangle=U_2|n\rangle
    \end{equation}
    where $n=0,1,2,\cdots,2^N-1$. Though more than one quantum states are offered, they are not prepared at the same state. Still, we are provided with no copies, which is the main difficulty. For given quantum states encoded by the same circuit, the structure shown in fig.(\ref{triangle}) can be used to find the encoding circuit. 
    
    
    As one can notice from the figure, we need in total $3N$ qubits in the circuit. The first $N$ qubits are prepared at the encoded states $|\psi(t)\rangle$, and the other $2N$ qubits are used as auxiliary qubits. 
    Note that the rotation gates $R_n$ satisfy $R_n|0\rangle=|n\rangle$. Initially, the system is prepared at $|\Psi_{in}\rangle=|\psi\rangle\otimes|0\rangle\otimes|0\rangle$, where every ket represents state of $N$ qubits. The very first step is to apply two control operations $S_1,S_2$, and 
    \begin{equation}
        S_1=\sum_{n=0}^{2^N-1}\left(|\psi_{1,n}\rangle\langle\psi_{1,n}|\otimes R_n\right)\otimes I
    \end{equation}
	\begin{equation}
        S_2=\sum_{n=0}^{2^N-1}\left(|\psi_{2,n}\rangle\langle\psi_{2,n}|\otimes I\otimes R_n\right)
    \end{equation}
    As these two operations commute with each other, it does not matter to change their order.
    
    Assume that the given qubits are at state $|\psi_{1,m}\rangle$, after these operations the system will be converted to
    \begin{equation}
        |\Psi_1\rangle=\sum_{n=0}^{2^N-1}\left[c^1_{m,n}|\psi_{2,n}\rangle\otimes|m\rangle\otimes|n\rangle\right]
    \end{equation}
	where $c^1_{m,n}=\langle\psi_{2,n}|\psi_{1,m}\rangle$. Now apply measurements on the auxiliary qubits so that the whole system will collapse. With possibility $P_{1,l}={|c^1_{m,l}|}^2$ one will find the state at
	\begin{equation}
	    |\Psi_{1,l}\rangle=|\psi_{2,l}\rangle\otimes|m\rangle\otimes|l\rangle
	\end{equation}
	Similarly, if the given qubits are prepared at state $|\psi_{2,m}\rangle$, then after these operations and measurements one will have possibility $P_{2,l}={|c^2_{m,l}|}^2$ to find the system at state
	\begin{equation}
	    |\Psi_{2,l}\rangle=|\psi_{1,l}\rangle\otimes|l\rangle\otimes|m\rangle
	\end{equation}
	and  $c^2_{m,n}=\langle\psi_{1,n}|\psi_{2,m}\rangle$.
	To distinguish $|\psi_{1,l}\rangle$ and $|\psi_{2,l}\rangle$, one convenient solution is to apply operation $U_1^\dagger$, and measure the final state. If the results is $|l\rangle$, then we can conclude that given states are encoded by $U_1$, otherwise we believe that they are encoded by $U_2$. Yet from one state the result might be not correct, as one might be led to wrong decision under two situations. Firstly, we could hardly move forward if $l=m$ after the first measurement. Besides, there is still chance $|\langle\psi_{1,l}|l\rangle|^2$ to get wrong result at the final step. As a conclusion, for a single given encoded state, possibility to derive the correct encoding circuit is 
	\begin{equation}
	    P_{1,m}=1-{|\langle\psi_{2,m}|\psi_{1,m}\rangle|}^2-\sum_{l=0,l\neq m}^{2^N-1}{\left[{|\langle\psi_{2,l}|\psi_{1,m}\rangle|}^2{|\langle\psi_{2,l}|l\rangle|}^2\right]}
	\end{equation}
	Further, as 
	\begin{equation}
	    \begin{split}
	         P_{1,m}&\geq1-{|\langle\psi_{2,m}|\psi_{1,m}\rangle|}^2-\left[\sum_{l=0,l\neq m}^{2^N-1}{{|\langle\psi_{2,l}|\psi_{1,m}\rangle|}^2}\right]{|\langle\psi_{2,l}|l\rangle|}^2\\
	         &=\left[1-{|\langle\psi_{2,m}|\psi_{1,m}\rangle|}^2\right]\left[1-{|\langle\psi_{2,l}|l\rangle|}^2\right]
	    \end{split}
	\end{equation}
	Once we can make sure that $\left[1-{|\langle\psi_{2,m}|\psi_{1,m}\rangle|}^2\right]\left[1-{|\langle\psi_{2,l}|l\rangle|}^2\right]\geq\frac{1}{2}$, then it would be a choice applying this method to find the encoding circuit. However, when more encoding operations are introduced, if would be much more difficult to find the encoding operation from one single state. In other words, decode the former character by find the encoding operation can not help Eve to wiretap the communication.
	
	\subsection*{Appropriate length of message}
	
	Intuitively, if the message contains infinite words, no matter how many previous qubits are included in the encoding process as shown in eq.(7), Eve will always be able to derive all following information, once she decodes enough characters continuously. Generally, the shorter the message is, the more difficult it will be for Eve to decode it. Since when designing the communication strategy, it will also be important to set an appropriate length. Here, we will show some numerical simulation results as an attempt to find the appropriate message length.
	
\begin{figure}[H]
	    \centering
        \includegraphics[width=0.45\linewidth]{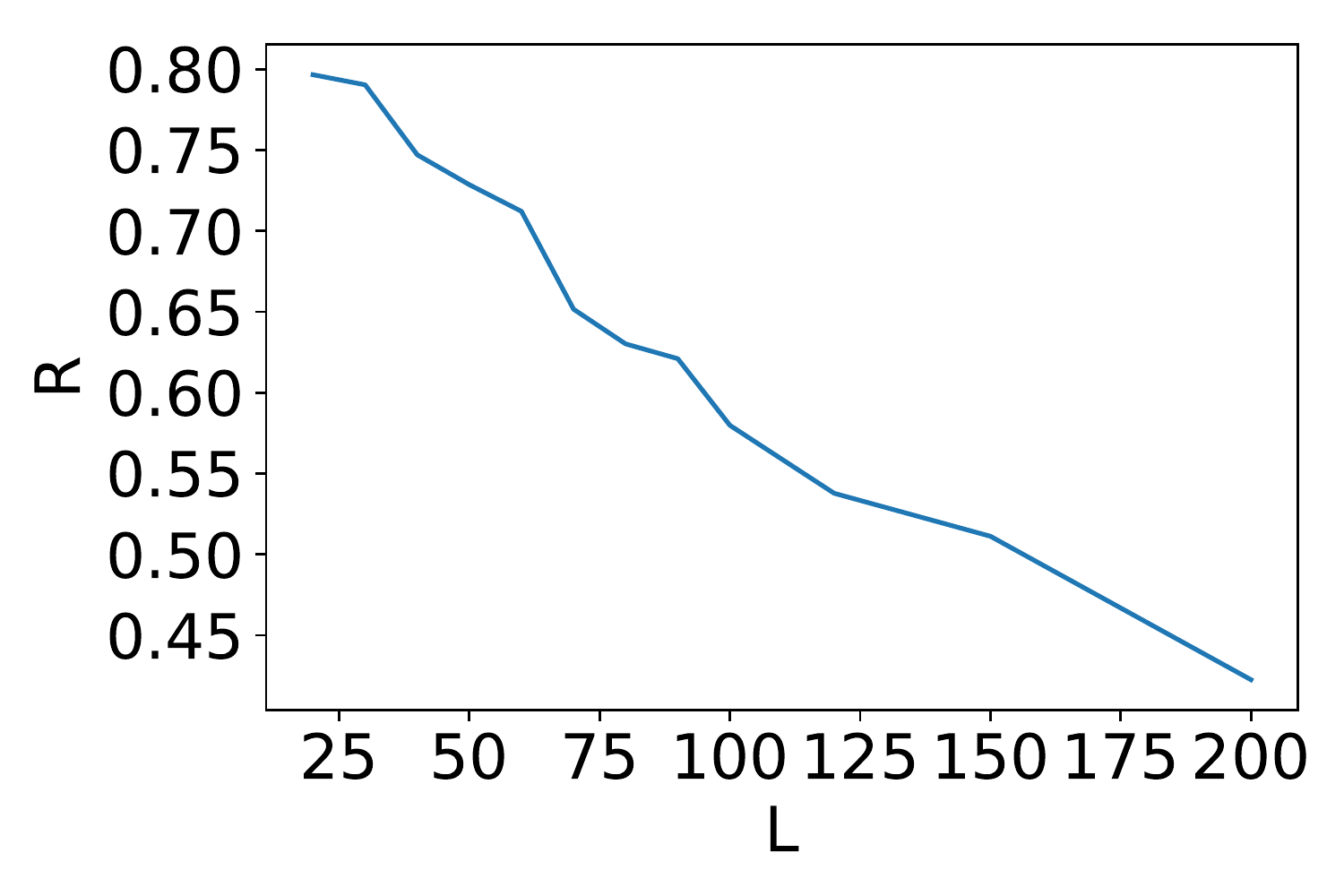}
	    \caption{{\bf Sketch of the trend of average err rate against the length of message.} Here {\bf L} represents the length of message and {\bf R} represents the average err rate. In the simulation, Alice and Bob transport message encoded by the improved strategy (As we discussed in Sec.3, two previous states contribute to the next encoding method). Eve can apply arbitrary operations on the qubits, and knows all details except the exact time $t=0$.}
	    \label{errrate}
	\end{figure}

	Fig.(\ref{errrate}) shows the trend of average err rate when the message becomes longer. The average err rate $R$ is defined as
	\begin{equation}
	    R=\frac{\sum_{j=1}^{j=M}e_j}{ML}
	\end{equation}
	where $M$ is the total times of simulation, and $L$ is length of the message, $e_j$ is the errs Eve makes in the $j$-th time when decoding the message.
	
\begin{figure}[H]
	    \centering
        \includegraphics[width=0.45\linewidth]{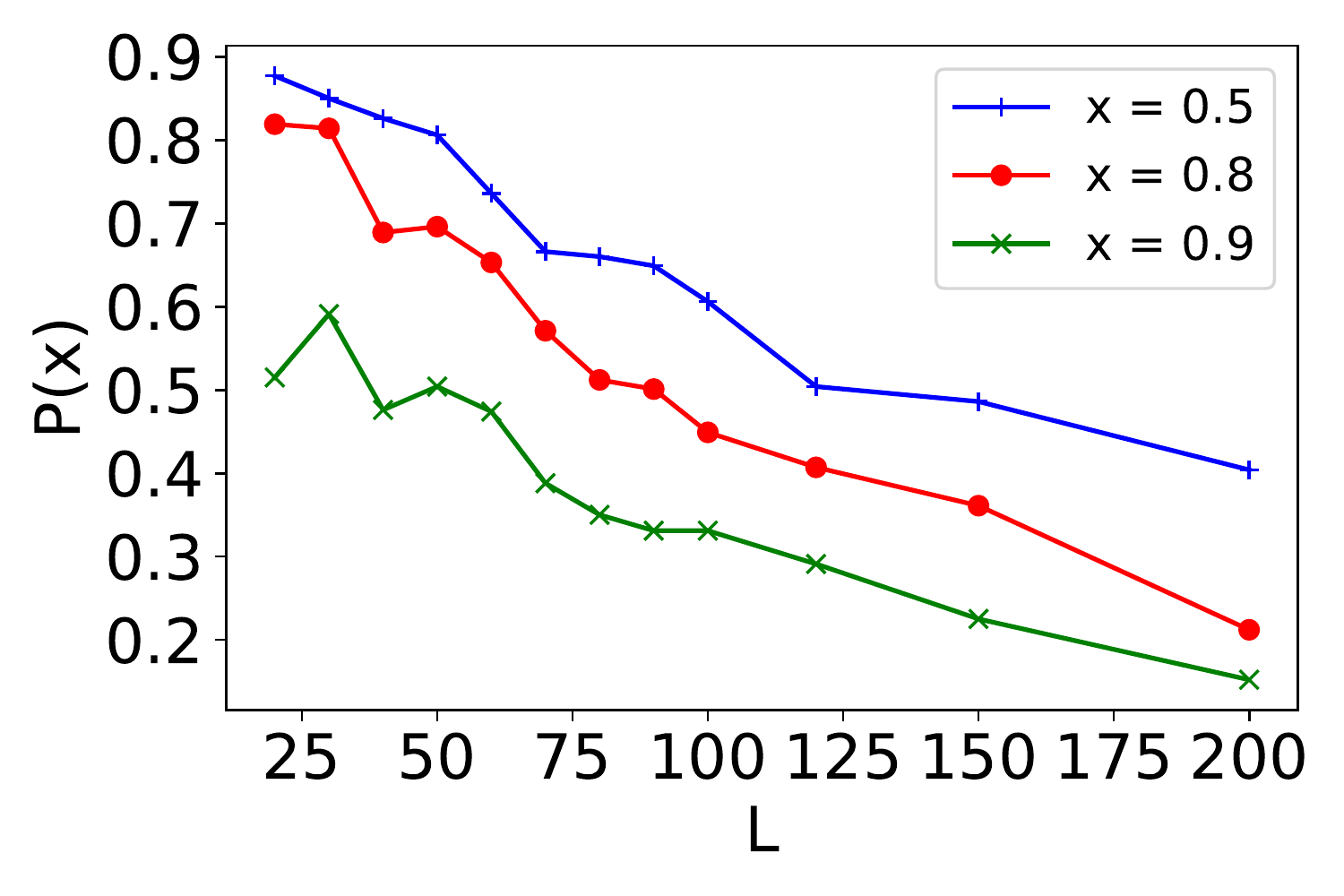}
	    \caption{{\bf Sketch of $P(x)$ against the length of message.} Here {\bf L} represents the length of message and {\bf P(x)} represents probability that Eve makes at least $xL$ mistakes when decoding the message. In the simulation, Alice and Bob transport message encoded by the improved strategy (As we discussed in Sec.3, two previous states contribute to the next encoding method). Eve can apply arbitrary operations on the qubits, and knows all details except the exact time $t=0$.}
	    \label{fig_px}
	\end{figure}

	Further, we also compare the performance of $P(x)$ under various length of message, which is the probability that Eve makes more than $xL$ mistakes when decoding a message with length $L$. Mathematically, $P(x)$ is defined as
	\begin{equation}
	    R=\frac{\sum_{j=1}^{j=M}\Gamma(e_j-xL)}{M}
	\end{equation}
	where $Gamma$ is used to represent the threshold function, and $Gamma(y)=1$ if $y>0$, otherwise $Gamma(y)=0$. Simulation results are shown in fig.(\ref{fig_px})
	
	Same as in the main article, here we set ${\bf \Theta_1}=(0.45\pi, 4.04, 1.04, 0.92)$ and ${\bf \Theta_2} = [0, 0.35, 0.55\pi, 0.79)$ (Just some random numbers, not the optimal ones). Based on the simulation we suggest that when only two previous states contribute to the encoding process, it is better to transport message around 100 characters or less once, otherwise the risk of wiretapping can no more be ignored. However, it can always be a solution to divide a long message into a branch of pieces.
	
	\subsection*{Encoding circuits for more qubits}
	Here we will expand the encoding circuits to more qubits. Generally, 6 qubits are enough for common communications based on letters and numbers. However, sometimes special characters might be required, and then more qubits should be included in the communication. Sill, the encoding operations will contain no more parameters except ${\bf \Theta_1, \Theta_2}$. Note that there are $N$ qubits $q_1, q_2, \dots, q_N$ used in communication.
	
	{\bf Encoding operations design method I:} A control gates loop that contains all qubits.
	
	Consider a permutation of the qubits, note the the location of $q_j$ as $p(j)$, where $0\le p(j)<N$, and $p(j)\neq p(k)$ when $j\neq k$. Then for every permutation $p$, there is a control gates loop as
	\begin{equation}
	    U(p) = \left[\prod_{j=0}^{N-1}u_{p_{j}p_{j+1}}\right]\cdot u_{p_{N}p_{0}}
	\end{equation}
	In the main article, we have defined that
    \begin{equation}
	    u_{ij}=|0\rangle\langle0|_i\otimes R_j({\bf \Theta_1})+ |1\rangle\langle 1|_i\otimes R_j({\bf \Theta_2})
	\end{equation}
	
	{\bf Encoding operations design method II:} Decompose the encoding operation as a combination of $U_{bi}$ and $U_{tri}$.
	
	If $N$ is even, we can always rewrite $N$ as a sum of $N/2$ qubit pairs. Further if $N\geq 6$, we can rewrite $N$ as a sum of some qubit triplets and some qubit pairs. For each qubit pair we can apply $U_{bi}$, and for each tripet we can apply $U_{tri}$. On the other hand, if $N\geq3$ is odd, we can rewrite $N=3+(N-3)$, where $N-3$ is now a even number. Consider that the $N$ qubits are divide into $a$ pairs and $b$ triplets, there exists a corresponding encoding operation,
	\begin{equation}
	    U=\bigotimes_{j=0}^{j=a}u_{bi}^j
	    \cdot\bigotimes_{k=0}^{k=b}u_{tri}^k
	\end{equation}
	where the superscript represents the $j-th$ qubit pair or the $k-th$ triplet. Note here the $u_{bi}$ is different from the $U_{bi}$ in the main article. Instead, $u_{bi}$ applies on qubit $q_i$ and $q_j$ is defined as
	\begin{equation}
	    u_{bi}=u_{ij}\cdot u_{ji}
	\end{equation}
	Similarly, $u_{tri}$ applies on qubit triplet $q_i,q_j,q_k$ is defined as 
	\begin{equation}
	    u_{tri}=u_{ij}\cdot u_{jk}\cdot u_{ki}
	\end{equation}

	\subsection*{Encoding operations}
    
    In this section we will offer the 64 encoding operations corresponding to different former words. To reduce confuse, here we use $V_n$ to represent the encoding operations. Subscript represents the former word, where each number represent a single letter, number or notation. As discussed in the main article, $0$ represents blank space, used as word divider. $1$ to $26$ represent characters capital letters 'A' to 'Z', $27$ to $52$ represent little letters 'a' to 'z', and $53$ to $62$ represent numbers '0', '1' to '9'. The last eigenstate $63$ represents ',', '.', or other marks to divide sentences. 
    As an example, $V_1$ will be used as the encoding operation at time $t$, if we find out that $n(t-1)=1$, in other words the former word is $'A'$.
    
    \begin{equation*}
        V_0 = (u_{12}\cdot u_{21})\otimes(u_{34}\cdot u_{43})\otimes(u_{56}\cdot u_{65})
    \end{equation*}
    
    \begin{equation*}
        V_{1}=(u_{12}\cdot u_{23} \cdot u_{31})\otimes (u_{45}\cdot u_{56} \cdot u_{64})
    \end{equation*}
    
    \begin{equation*}
        V_{2}=(u_{12}\cdot u_{23} \cdot u_{31})\otimes (u_{46}\cdot u_{65} \cdot u_{54})
    \end{equation*}
    
    \begin{equation*}
        V_{3}=(u_{12}\cdot u_{23} \cdot u_{31})\otimes (u_{54}\cdot u_{46} \cdot u_{65})
    \end{equation*}
    
    \begin{equation*}
        V_{4}=(u_{12}\cdot u_{23} \cdot u_{31})\otimes (u_{56}\cdot u_{64} \cdot u_{45})
    \end{equation*}
    
    \begin{equation*}
        V_{5}=(u_{12}\cdot u_{23} \cdot u_{31})\otimes (u_{64}\cdot u_{45} \cdot u_{56})
    \end{equation*}
    
    \begin{equation*}
        V_{6}=(u_{12}\cdot u_{23} \cdot u_{31})\otimes (u_{65}\cdot u_{54} \cdot u_{46})
    \end{equation*}
    
    \begin{equation*}
        V_{7}=(u_{13}\cdot u_{32} \cdot u_{21})\otimes (u_{45}\cdot u_{56} \cdot u_{64})
    \end{equation*}
    
    \begin{equation*}
        V_{8}=(u_{13}\cdot u_{32} \cdot u_{21})\otimes (u_{46}\cdot u_{65} \cdot u_{54})
    \end{equation*}
    
    \begin{equation*}
        V_{9}=(u_{13}\cdot u_{32} \cdot u_{21})\otimes (u_{54}\cdot u_{46} \cdot u_{65})
    \end{equation*}
    
    \begin{equation*}
        V_{10}=(u_{13}\cdot u_{32} \cdot u_{21})\otimes (u_{56}\cdot u_{64} \cdot u_{45})
    \end{equation*}
    
    \begin{equation*}
        V_{11}=(u_{13}\cdot u_{32} \cdot u_{21})\otimes (u_{64}\cdot u_{45} \cdot u_{56})
    \end{equation*}
    
    \begin{equation*}
        V_{12}=(u_{13}\cdot u_{32} \cdot u_{21})\otimes (u_{65}\cdot u_{54} \cdot u_{46})
    \end{equation*}
    
    \begin{equation*}
        V_{13}=(u_{21}\cdot u_{13} \cdot u_{32})\otimes (u_{45}\cdot u_{56} \cdot u_{64})
    \end{equation*}
    
    \begin{equation*}
        V_{14}=(u_{21}\cdot u_{13} \cdot u_{32})\otimes (u_{46}\cdot u_{65} \cdot u_{54})
    \end{equation*}
    
    \begin{equation*}
        V_{15}=(u_{21}\cdot u_{13} \cdot u_{32})\otimes (u_{54}\cdot u_{46} \cdot u_{65})
    \end{equation*}
    
    \begin{equation*}
        V_{16}=(u_{21}\cdot u_{13} \cdot u_{32})\otimes (u_{56}\cdot u_{64} \cdot u_{45})
    \end{equation*}
    
    \begin{equation*}
        V_{17}=(u_{21}\cdot u_{13} \cdot u_{32})\otimes (u_{64}\cdot u_{45} \cdot u_{56})
    \end{equation*}
    
    \begin{equation*}
        V_{18}=(u_{21}\cdot u_{13} \cdot u_{32})\otimes (u_{65}\cdot u_{54} \cdot u_{46})
    \end{equation*}
    
    \begin{equation*}
        V_{19}=(u_{23}\cdot u_{31} \cdot u_{12})\otimes (u_{45}\cdot u_{56} \cdot u_{64})
    \end{equation*}
    
    \begin{equation*}
        V_{20}=(u_{23}\cdot u_{31} \cdot u_{12})\otimes (u_{46}\cdot u_{65} \cdot u_{54})
    \end{equation*}
    
    \begin{equation*}
        V_{21}=(u_{23}\cdot u_{31} \cdot u_{12})\otimes (u_{54}\cdot u_{46} \cdot u_{65})
    \end{equation*}
    
    \begin{equation*}
        V_{22}=(u_{23}\cdot u_{31} \cdot u_{12})\otimes (u_{56}\cdot u_{64} \cdot u_{45})
    \end{equation*}
    
    \begin{equation*}
        V_{23}=(u_{23}\cdot u_{31} \cdot u_{12})\otimes (u_{64}\cdot u_{45} \cdot u_{56})
    \end{equation*}
    
    \begin{equation*}
        V_{24}=(u_{23}\cdot u_{31} \cdot u_{12})\otimes (u_{65}\cdot u_{54} \cdot u_{46})
    \end{equation*}
    
    \begin{equation*}
        V_{25}=(u_{31}\cdot u_{12} \cdot u_{23})\otimes (u_{45}\cdot u_{56} \cdot u_{64})
    \end{equation*}
    
    \begin{equation*}
        V_{26}=(u_{31}\cdot u_{12} \cdot u_{23})\otimes (u_{46}\cdot u_{65} \cdot u_{54})
    \end{equation*}
    
    \begin{equation*}
        V_{27}=(u_{13}\cdot u_{35} \cdot u_{51})\otimes (u_{24}\cdot u_{46} \cdot u_{62})
    \end{equation*}
    
    \begin{equation*}
        V_{28}=(u_{13}\cdot u_{35} \cdot u_{51})\otimes (u_{26}\cdot u_{64} \cdot u_{42})
    \end{equation*}
    
    \begin{equation*}
        V_{29}=(u_{13}\cdot u_{35} \cdot u_{51})\otimes (u_{42}\cdot u_{26} \cdot u_{64})
    \end{equation*}
    
    \begin{equation*}
        V_{30}=(u_{13}\cdot u_{35} \cdot u_{51})\otimes (u_{46}\cdot u_{62} \cdot u_{24})
    \end{equation*}
    
    \begin{equation*}
        V_{31}=(u_{13}\cdot u_{35} \cdot u_{51})\otimes (u_{62}\cdot u_{24} \cdot u_{46})
    \end{equation*}
    
    \begin{equation*}
        V_{32}=(u_{13}\cdot u_{35} \cdot u_{51})\otimes (u_{64}\cdot u_{42} \cdot u_{26})
    \end{equation*}
    
    \begin{equation*}
        V_{33}=(u_{15}\cdot u_{53} \cdot u_{31})\otimes (u_{24}\cdot u_{46} \cdot u_{62})
    \end{equation*}
    
    \begin{equation*}
        V_{34}=(u_{15}\cdot u_{53} \cdot u_{31})\otimes (u_{26}\cdot u_{64} \cdot u_{42})
    \end{equation*}
    
    \begin{equation*}
        V_{35}=(u_{15}\cdot u_{53} \cdot u_{31})\otimes (u_{42}\cdot u_{26} \cdot u_{64})
    \end{equation*}
    
    \begin{equation*}
        V_{36}=(u_{15}\cdot u_{53} \cdot u_{31})\otimes (u_{46}\cdot u_{62} \cdot u_{24})
    \end{equation*}
    
    \begin{equation*}
        V_{37}=(u_{15}\cdot u_{53} \cdot u_{31})\otimes (u_{62}\cdot u_{24} \cdot u_{46})
    \end{equation*}
    
    \begin{equation*}
        V_{38}=(u_{15}\cdot u_{53} \cdot u_{31})\otimes (u_{64}\cdot u_{42} \cdot u_{26})
    \end{equation*}
    
    \begin{equation*}
        V_{39}=(u_{31}\cdot u_{15} \cdot u_{53})\otimes (u_{24}\cdot u_{46} \cdot u_{62})
    \end{equation*}
    
    \begin{equation*}
        V_{40}=(u_{31}\cdot u_{15} \cdot u_{53})\otimes (u_{26}\cdot u_{64} \cdot u_{42})
    \end{equation*}
    
    \begin{equation*}
        V_{41}=(u_{31}\cdot u_{15} \cdot u_{53})\otimes (u_{42}\cdot u_{26} \cdot u_{64})
    \end{equation*}
    
    \begin{equation*}
        V_{42}=(u_{31}\cdot u_{15} \cdot u_{53})\otimes (u_{46}\cdot u_{62} \cdot u_{24})
    \end{equation*}
    
    \begin{equation*}
        V_{43}=(u_{31}\cdot u_{15} \cdot u_{53})\otimes (u_{62}\cdot u_{24} \cdot u_{46})
    \end{equation*}
    
    \begin{equation*}
        V_{44}=(u_{31}\cdot u_{15} \cdot u_{53})\otimes (u_{64}\cdot u_{42} \cdot u_{26})
    \end{equation*}
    
    \begin{equation*}
        V_{45}=(u_{35}\cdot u_{51} \cdot u_{13})\otimes (u_{24}\cdot u_{46} \cdot u_{62})
    \end{equation*}
    
    \begin{equation*}
        V_{46}=(u_{35}\cdot u_{51} \cdot u_{13})\otimes (u_{26}\cdot u_{64} \cdot u_{42})
    \end{equation*}
    
    \begin{equation*}
        V_{47}=(u_{35}\cdot u_{51} \cdot u_{13})\otimes (u_{42}\cdot u_{26} \cdot u_{64})
    \end{equation*}
    
    \begin{equation*}
        V_{48}=(u_{35}\cdot u_{51} \cdot u_{13})\otimes (u_{46}\cdot u_{62} \cdot u_{24})
    \end{equation*}
    
    \begin{equation*}
        V_{49}=(u_{35}\cdot u_{51} \cdot u_{13})\otimes (u_{62}\cdot u_{24} \cdot u_{46})
    \end{equation*}
    
    \begin{equation*}
        V_{50}=(u_{35}\cdot u_{51} \cdot u_{13})\otimes (u_{64}\cdot u_{42} \cdot u_{26})
    \end{equation*}
    
    \begin{equation*}
        V_{51}=(u_{51}\cdot u_{13} \cdot u_{35})\otimes (u_{24}\cdot u_{46} \cdot u_{62})
    \end{equation*}
    
    \begin{equation*}
        V_{52}=(u_{51}\cdot u_{13} \cdot u_{35})\otimes (u_{26}\cdot u_{64} \cdot u_{42})
    \end{equation*}
    
    \begin{equation*}
        V_{53}=(u_{51}\cdot u_{13} \cdot u_{35})\otimes (u_{42}\cdot u_{26} \cdot u_{64})
    \end{equation*}
    
    \begin{equation*}
        V_{54}=(u_{51}\cdot u_{13} \cdot u_{35})\otimes (u_{46}\cdot u_{62} \cdot u_{24})
    \end{equation*}
    
    \begin{equation*}
        V_{55}=(u_{51}\cdot u_{13} \cdot u_{35})\otimes (u_{62}\cdot u_{24} \cdot u_{46})
    \end{equation*}
    
    \begin{equation*}
        V_{56}=(u_{51}\cdot u_{13} \cdot u_{35})\otimes (u_{64}\cdot u_{42} \cdot u_{26})
    \end{equation*}
    
    \begin{equation*}
        V_{57}=(u_{53}\cdot u_{31} \cdot u_{15})\otimes (u_{24}\cdot u_{46} \cdot u_{62})
    \end{equation*}
    
    \begin{equation*}
        V_{58}=(u_{53}\cdot u_{31} \cdot u_{15})\otimes (u_{26}\cdot u_{64} \cdot u_{42})
    \end{equation*}
    
    \begin{equation*}
        V_{59}=(u_{53}\cdot u_{31} \cdot u_{15})\otimes (u_{42}\cdot u_{26} \cdot u_{64})
    \end{equation*}
    
    \begin{equation*}
        V_{60}=(u_{53}\cdot u_{31} \cdot u_{15})\otimes (u_{46}\cdot u_{62} \cdot u_{24})
    \end{equation*}
    
    \begin{equation*}
        V_{61}=(u_{53}\cdot u_{31} \cdot u_{15})\otimes (u_{62}\cdot u_{24} \cdot u_{46})
    \end{equation*}
    
    \begin{equation*}
        V_{62}=(u_{53}\cdot u_{31} \cdot u_{15})\otimes (u_{64}\cdot u_{42} \cdot u_{26})
    \end{equation*}
    
    \begin{equation*}
        V_{63} = (u_{21}\cdot u_{12})\otimes(u_{43}\cdot u_{34})\otimes(u_{65}\cdot u_{56})
    \end{equation*}
    
    Please note that the above is just one design of the encoding operations. One can design arbitrary encoding sets for various demands.

\end{document}